\documentclass[traditabstract]{aa}

\usepackage{graphicx}
\usepackage{txfonts}
\usepackage{natbib}
\usepackage{xcolor}
\usepackage{xspace}
\usepackage[colorlinks=true,linkcolor=blue,citecolor=blue]{hyperref}
\usepackage{etoolbox}

\graphicspath{{.}}

\makeatletter

\patchcmd{\NAT@citex}
  {\@citea\NAT@hyper@{%
     \NAT@nmfmt{\NAT@nm}%
     \hyper@natlinkbreak{\NAT@aysep\NAT@spacechar}{\@citeb\@extra@b@citeb}%
     \NAT@date}}
  {\@citea\NAT@nmfmt{\NAT@nm}%
   \NAT@aysep\NAT@spacechar\NAT@hyper@{\NAT@date}}{}{}

\patchcmd{\NAT@citex}
  {\@citea\NAT@hyper@{%
     \NAT@nmfmt{\NAT@nm}%
     \hyper@natlinkbreak{\NAT@spacechar\NAT@@open\if*#1*\else#1\NAT@spacechar\fi}%
       {\@citeb\@extra@b@citeb}%
     \NAT@date}}
  {\@citea\NAT@nmfmt{\NAT@nm}%
   \NAT@spacechar\NAT@@open\if*#1*\else#1\NAT@spacechar\fi\NAT@hyper@{\NAT@date}}
  {}{}

\makeatother

\newcommand{\hlink}[1]{\url{http://#1}\xspace}

\newcommand{\rfig}[1]{Fig.~\ref{#1}}
\newcommand{\rfigs}[1]{Figs.~\ref{#1}}
\newcommand{\req}[1]{Eq.~\ref{#1}}
\newcommand{\reqs}[1]{Eqs.~\ref{#1}}
\newcommand{\rtab}[1]{Table \ref{#1}}
\newcommand{\rtabs}[1]{Tables \ref{#1}}

\newcommand{\rsec}[1]{section \ref{#1}}
\newcommand{\rsecs}[1]{sections \ref{#1}}

\newcommand{\herschel}{{\it Herschel}\xspace}
\newcommand{\spitzer}{{\it Spitzer}\xspace}

\newcommand{\hst}{{\it HST}\xspace}
\newcommand{\jwst}{{\it JWST}\xspace}

\newcommand{\um}{\mu{\rm m}}
\newcommand{\mm}{{\rm mm}}
\newcommand{\uJy}{\mu{\rm Jy}}
\newcommand{\mJy}{{\rm mJy}}

\newcommand{\sfe}{{\rm SFE}}
\newcommand{\sfr}{{\rm SFR}}

\newcommand{\sfrms}{{\rm SFR}_{\rm MS}}
\newcommand{\ssfr}{{\rm sSFR}}
\newcommand{\lir}{L_{\rm IR}}

\newcommand{\leight}{L_8}
\newcommand{\ireight}{{\rm IR8}}

\newcommand{\luv}{L_{\rm UV}}
\newcommand{\lsun}{L_\odot}
\newcommand{\msun}{{\rm M}_\odot}
\newcommand{\mgas}{M_{\rm gas}}

\newcommand{\dex}{{\rm dex}}
\newcommand{\mstar}{M_\ast}

\newcommand{\tdust}{T_{\rm dust}}

\newcommand{\rsb}{R_{\rm SB}}
\newcommand{\uvj}{$UVJ$\xspace}

\newcommand{\mean}[1]{\left<#1\right>}
\newcommand{\fpah}{f_{\rm PAH}}
\newcommand{\mdust}{M_{\rm dust}}
\newcommand{\kelvin}{{\rm K}}

\newcommand{\Ks}{$K_{\rm s}$\xspace}

\defcitealias{schreiber2015}{S15}
\defcitealias{chary2001}{CE01}
\defcitealias{dale2002}{DH02}
\defcitealias{galliano2011}{G11}
\defcitealias{draine2007}{DL07}

\newcommand*\dd{\ensuremath{d}}

\begin{document}

\title{Dust temperature and mid-to-total infrared color \\ distributions for star-forming galaxies at $0<z<4$\thanks{The dust library described in this paper is available publicly at \url{http://cschreib.github.io/s17-irlib/}} \thanks{\rtabs{TAB:lconv_alma} to \ref{TAB:lconv} are only available in electronic form
at the CDS via anonymous ftp to \url{cdsarc.u-strasbg.fr} \texttt{(130.79.128.5)}
or via \url{http://cdsweb.u-strasbg.fr/cgi-bin/qcat?J/A+A/x/y}}}

\author{C.~Schreiber\inst{1,2}
  \and D.~Elbaz\inst{2}
  \and M.~Pannella\inst{3}
  \and L.~Ciesla\inst{2}
  \and T.~Wang\inst{4,2}
  \and M.~Franco\inst{2}
}

\institute{
    Leiden Observatory, Leiden University, NL-2300 RA Leiden, The Netherlands \\
    \email{cschreib@strw.leidenuniv.nl}
    \and
    Laboratoire AIM-Paris-Saclay, CEA/DSM/Irfu - CNRS - Universit\'e Paris Diderot, CEA-Saclay, pt courrier 131, F-91191 Gif-sur-Yvette, France
    \and Faculty of Physics, Ludwig-Maximilians Universit\"at, Scheinerstr.\ 1, 81679 Munich, Germany
    \and
    School of Astronomy and Astrophysics, Nanjing University, Nanjing, 210093, China
}

\date{Received 4th of July 2017; accepted 24th of October 2017}

\abstract {
We present a new, publicly available library of dust spectral energy distributions (SEDs). These SEDs are characterized by only three parameters: the dust mass ($\mdust$), the dust temperature ($\tdust$), and the mid-to-total infrared color ($\ireight\equiv\lir/\leight$). The latter measures the relative contribution of polycyclic aromatic hydrocarbon (PAH) molecules to the total infrared luminosity. We used this library to model star-forming galaxies at $0.5<z<4$ in the deep CANDELS fields, using both individual detections and stacks of \herschel and ALMA imaging, and extending this sample to $z=0$ using the \herschel Reference Survey. At first order, the dust SED of a galaxy was observed to be independent of stellar mass, but evolving with redshift. We found trends of increasing $\tdust$ and $\ireight$ with redshift and distance from the $\sfr$--$\mstar$ main sequence, and quantified for the first time their intrinsic scatter. Half of the observed variations of these parameters was captured by the above empirical relations, and after subtracting the measurement errors we found residual scatters of $\Delta\tdust/\tdust=12\%$ and $\Delta\log\ireight=0.18\,\dex$. We observed second order variations with stellar mass: massive galaxies ($\mstar > 10^{11}\,\msun$) at $z\le1$ have slightly lower temperatures indicative of a reduced star formation efficiency, while low mass galaxies ($\mstar < 10^{10}\,\msun$) at $z\ge1$ showed reduced PAH emission, possibly linked to their lower metallicities. Building on these results, we constructed high-fidelity mock galaxy catalogs to predict the accuracy of infrared luminosities and dust masses determined using a single broadband measurement. Using a single {\it James Webb Space Telescope} (\jwst) MIRI band, we found that $\lir$ is typically uncertain by $0.15\,\dex$, with a maximum of $0.25\,\dex$ when probing the rest-frame $8\,\um$, and this is not significantly impacted by typical redshift uncertainties. On the other hand, we found that ALMA bands 8 to 7 and 6 to 3 measured the dust mass at better than $0.2$ and $0.15\,\dex$, respectively, and independently of redshift, while bands 9 to 6 only measured $\lir$ at better than $0.2\,\dex$ at $z>1$, $3.2$, $3.8$, and $5.7$, respectively. Starburst galaxies had their $\lir$ significantly underestimated when measured by a single \jwst or ALMA band, while their dust mass from a single ALMA band were moderately overestimated. This dust library and the results of this paper can be used immediately to improve the design of observing proposals, and interpret more accurately the large amount of archival data from \spitzer, \herschel and ALMA.
}


\maketitle

\section{Introduction \label{SEC:introduction}}

Properly accounting for the amount of stellar light absorbed by dust has proven to be a key ingredient to study star formation in galaxies. The most obvious breakthrough linked to deep infrared (IR) surveys was probably the exciting new outlook they provided on the cosmic history of star formation \citep[e.g.,][and references therein]{smail1997,hughes1998,barger1998,blain1999,elbaz1999,flores1999,lagache1999,gispert2000,franceschini2001,elbaz2002,papovich2004,lefloch2005,elbaz2007,daddi2009-a,magnelli2009,gruppioni2010,elbaz2011,rodighiero2011,magdis2012,madau2014}. In the meantime, the emission of dust in distant galaxies has also been used to study the dust itself, which turned out to be a valuable tool to learn more about non-stellar baryonic matter, present in the interstellar medium (ISM) either in the form of dust grains or atomic and molecular gas \citep[e.g.,][]{chapman2003,hwang2010-a,elbaz2011,magdis2012,berta2013,scoville2014,santini2014,bethermin2015-a,genzel2015,tacconi2017}.

In the Local Universe, the large amount of infrared data acquired in the Milky Way and nearby galaxies has given birth to detailed models aiming to provide a description of the dust content from first principles \citep[e.g.,][to only name a few of the most recent ones]{zubko2004,draine2007,galliano2011,jones2013}. These models typically contain three main components \citep[see, e.g.,][]{desert1990}: big grains (BGs, $>0.01\,\um$), very small grains (VSGs, $<0.01\,\um$), and complex molecules (polycyclic aromatic hydrocarbon, or PAH). The most prominent one is the emission of big grains, which are at thermal equilibrium with the ambient interstellar radiation. These grains radiate like gray bodies with a typical temperature of $\tdust \sim 20$--$40\kelvin$, and therefore emit the bulk of their energy in the far-infrared (FIR) around the rest-frame $100\,\um$. Smaller grains have a too small cross-section to be at equilibrium with the ambient radiation, and are instead only transiently heated to temperatures $\sim1000\,\kelvin$. This produces continuum emission in the mid-infrared (MIR). Lastly, PAHs are large carbonated molecules which cool down through numerous rotational and vibrational modes, and thus produce a group of bright and broad emission lines between $\lambda=3.3$ and $12.3\,\um$ \citep{leger1984,allamandola1985}.

To reproduce a set of observations in the IR, one can vary the total mass of dust encompassing all ISM components ($\mdust$), the distribution of energy they receive from their surrounding medium ($U$), coming mostly from stellar light, and the properties of each species of grains and molecules, including their size distribution, their chemical state and composition (neutral vs.~ionized PAHs, silicate vs.~carbonated grains). Given these parameters, models can output the expected infrared spectrum and interpret the observed data. However, most of these parameters are degenerate or unconstrained and the number of degrees of freedom is too large, hence assumptions have to be made when applying such models to photometric data. Typical approaches (e.g., \citealt{draine2007}, hereafter \citetalias{draine2007} or \citealt{dacunha2008}) assume fixed grain distributions (motivated by observations from clouds of the Milky Way), and consider simplified geometries of dusty regions (e.g., birth clouds, diffuse ISM, hot torus around a super-massive black hole). This still allows much flexibility in the output spectrum, and can describe observations accurately (e.g., \citealt{dacunha2015,gobat2017}).

But even then, properly constraining the fit parameters (in particular the dust temperature) requires exquisite IR spectral energy distributions (SEDs) with good wavelength sampling, at a level of quality that can currently only be reached either in the Local Universe or at high-redshifts for the most extreme starbursts \citep[e.g.,][]{hwang2010-a,magdis2010-c,riechers2013}, strongly lensed galaxies \citep[e.g.,]{sklias2014}, or on stacked samples \citep[e.g.,][]{magnelli2014,bethermin2015-a}. For the typical higher redshift galaxy, the available IR SED is limited to one or two photometric points \citep[e.g.,][]{elbaz2011}, and even simpler approaches are often preferred. A number of empirical libraries have been constructed to this end, each composed of a reduced number of template SEDs. These templates are typically associated to different values of a single parameter, for example the $8$-to-$1000\,\um$ luminosity ($\lir$) \citep[][hereafter \citetalias{chary2001}]{chary2001}, a FIR color \citep[][hearafter \citetalias{dale2002}]{dale2002}, or the average intensity of the interstellar radiation field, $\mean{U}$ \citep{magdis2012,bethermin2015-a}. In all cases the number of free parameters is reduced to two: the normalization of the template (which can be linked either to the mass or luminosity of the dust) and its ``shape'' (essentially its average dust temperature, which defines the wavelength at which the template peaks). Despite their extreme simplicity, these models are sufficient to reproduce the observed IR features of the vast majority of distant galaxies, illustrating the fact that the dust SED of a galaxy taken as a whole is close to universal \citep[see, e.g.,][]{elbaz2010,elbaz2011}.

This universality echoes another key observation of the last decade: the main sequence of star-forming galaxies \citep{noeske2007-a,elbaz2007}. This tight correlation between the star formation rate ($\sfr$) and the stellar mass ($\mstar$) has been observed across a broad range of redshifts up to $z\sim6$ \citep[e.g.,][]{daddi2007-a,pannella2009-a,rodighiero2011,whitaker2012-a,bouwens2012,whitaker2014,salmon2015,pannella2015,schreiber2015,schreiber2017-a}. Since its discovery, the main sequence has been used to put upper limits on the variability of the star formation histories, showing that galaxies form their stars mostly through a unique and secular way, as opposed to random bursts \citep[see, e.g., the discussion in][]{noeske2007-a}.

Through observations of their molecular gas content, galaxies belonging to the main sequence have also been shown to form their stars with a roughly constant efficiency ($\sfe\equiv\sfr/\mgas$) (e.g., \citealt{daddi2010-a,tacconi2013,genzel2015}). The same conclusion can be drawn from the dust emission of these galaxies and the measurement of their average dust temperature (\citealt{magdis2012,bethermin2015-a}, but see however \citealt{schreiber2016}). Indeed, $\tdust$ is a proxy for $\lir/\mdust$, which itself can be linked directly to $\sfe/Z$ \citep{magdis2012}, where $Z$ is the gas-phase metallicity. The universality of the dust SED therefore also suggests that star formation in galaxies is the product of a universal mechanism, which still remains to be fully understood (see, e.g., \citealt{dekel2013} or \citealt{tacchella2015}).

Departures from this ``universal'' SED do exist however. Galaxy-to-galaxy variations of $\tdust$ have been observed, with a first correlation identified with $\lir$ \citep[e.g.,][]{soifer1987,soifer1989,dunne2000,chapman2003,chapin2009,symeonidis2009,amblard2010,hwang2010-a}. It was later argued that this correlation is not fundamental, but in fact consequential of two effects: on the one hand a global increase of the temperature with redshift \citep[e.g.,][]{magdis2012,magnelli2014,bethermin2015-a}, and on the other hand an additional increase of temperature for galaxies that are offset from the main sequence \citep{elbaz2011,magnelli2014,bethermin2015-a}, suggesting these galaxies form stars more efficiently than the average. Quantifying changes of the dust temperature can thus provide crucial information about the star formation efficiency in galaxies, and it is therefore an important ingredient in any library.

In addition, significant galaxy-to-galaxy variations have been observed in the MIR around the rest-frame $3$ to $12\,\um$. As written above, the dust emission in this wavelength domain is mostly produced by small grains and PAHs. The PAH emission lines are so bright that they typically contribute about $80\%$ of the observed broadband MIR fluxes \citep[e.g.,][]{helou2000,huang2007}, however their strength is strongly reduced in starbursts and active galactic nuclei (AGNs, e.g., \citep{armus2007}), in which hot dust takes over. Therefore, the observed diversity in the MIR can be expected to come mostly from a diversity of PAH properties, at least for galaxies without strong AGNs (e.g., \citealt{fritz2006}). The interplay between the overall strength of PAHs and physical conditions inside the host galaxy is not yet fully understood. Two main trends are known at present: on the one hand an anti-correlation with metallicity \citep[e.g.,][]{madden2006,wu2006,ohalloran2006,smith2007,draine2007-a,galliano2008,ciesla2014,remy-ruyer2015}, and on the other hand a correlation with $\lir$ \citep[e.g.,][]{pope2008-a,elbaz2011,nordon2012}. Although this latter correlation suffers from a significant scatter, it implies that PAH features or the $8\,\um$ luminosity can be used as a rough tracer of star formation rate \citep{pope2008-a,shipley2016}.

An interesting property of PAHs is that they are set aglow mostly in photo-dissociation regions, at the interface between the ionized and molecular interstellar medium \citep[e.g.,][]{tielens1993}, whereas the FIR dust continuum is emitted from the whole volume of the dust clouds. Therefore, by relating the dust continuum to the PAH emission one can probe the geometry of star-forming regions, and in particular the filling factor of \ion{H}{ii} regions. Using this approach and combining \spitzer and \herschel data, \cite{elbaz2011} have used the $\ireight=\lir/\leight$ ratio as a tracer of compactness in distant galaxies: at fixed $\lir$, a lower $\leight$ indicates a higher filling factor of \ion{H}{ii} regions, hence a higher compactness. main sequence galaxies have a constant $\ireight\sim4$, while, as for the dust temperature, $\ireight$ increases as a function of the distance to the main sequence \citep[see also][]{nordon2012,rujopakarn2013,murata2014}. These trends confirm that galaxies above the main sequence form their stars in a different way, with a higher efficiency and in more compact volumes.

While the study of the physical origin of the MIR emission is obviously of interest on its own, it is also important to the extra-galactic community for practical observational reasons. Since the PAH emission is strong and found at low infrared wavelengths, the wavelength domain around the rest-frame $8\,\um$ is easier to observe than than the FIR continuum. It is particularly the case for galaxies at $z\sim2$, where the rest-frame $8\,\um$ shifts into the very deep \spitzer MIPS $24\,\um$ band, and allows the detection of galaxies significantly fainter than the detection limit of other infrared observatories like \herschel. However, these galaxies have by construction a very poorly constrained infrared SED, and extrapolating the total $\lir$ from the $8\,\um$ alone is challenging \citep[see ][]{daddi2007-a,elbaz2011,magdis2011-a,rujopakarn2013,shivaei2017}. Doing so requires an accurate understanding of the $\ireight$ ratio.

Another important practical interest for the rest-frame $8\,\um$ is that it will be easily accessed by the {\it James Webb Space Telescope} (\jwst) in the near future, for both local and distant galaxies. Once this satellite is launched, there will be a need for a properly calibrated library to exploit these data together with ancillary \herschel and \spitzer observations, and in particular to cope with their absence for the faintest objects.

Our goal in this paper is the following. We introduce in \rsec{SEC:irsed} a new SED library in which both $\tdust$ and $\ireight$ are free parameters. This library provides an increased level of detail compared to standard libraries (e.g., \citetalias{chary2001}, \citetalias{dale2002}), but still keeps the number of adjustable parameters low. In \rsec{SEC:irsed_stack} we determine the redshift evolution of both $\tdust$ and $\ireight$ using the MIR-to-FIR stacks introduced in \cite{schreiber2015}, to which we add stacks of $16\,\um$ and ALMA $870\,\um$ to better constrain the PAH features and the dust temperature at high redshifts. We then apply this model to individual \herschel detections in \rsec{SEC:irsed_indiv} to constrain the scatter on the model parameters, and also to quantify their enhancements for those galaxies that are offset from the main sequence. Using these results, we derive in \rsec{SEC:irsed_recipe} a set of recipes for optimal SED fitting in the IR, in particular when a single photometric band is available. Finally, we quantify the accuracy of such measurements using mock galaxy catalogs in \rsec{SEC:irsed_mono}, and provide in \rsec{SEC:conv} conversion factors to determine dust masses and infrared luminosities from ALMA fluxes and \jwst MIRI luminosities. These are valid for $0<z<4$, and are extrapolated to $z=8$ for ALMA.

In the following, we assume a $\Lambda$CDM cosmology with $H_0 = 70\ {\rm km}\,{\rm s}^{-1} {\rm Mpc}^{-1}$, $\Omega_{\rm M} = 0.3$, $\Omega_\Lambda = 0.7$ and a \cite{salpeter1955} initial mass function (IMF), to derive both star formation rates and stellar masses. All magnitudes are quoted in the AB system, such that $M_{\rm AB} = 23.9 - 2.5\log_{10}(S_{\!\nu}\ [\uJy])$.

\section{Sample and observations}

We based this analysis on the sample and data described in \cite{schreiber2015} \citepalias[hereafter][]{schreiber2015}, which covers redshifts from $z=0.3$ to $z=4$. We complemented this sample with $z=0$ galaxies from the \herschel Reference Survey (HRS; \citealt{boselli2010}), and $z=2$ to $4$ galaxies in the Extended Chandra Deep Field South (ECDFS) observed by ALMA as part of the ALESS program \citep{hodge2013}. In this section, we make a brief summary of these observations.

\subsection{CANDELS}

The catalogs we used in this work are based on the CANDELS \citep{grogin2011,koekemoer2011} \emph{Hubble Space Telescope} (\hst) WFC3 $H$ band images in the fields covered by deep \herschel PACS and SPIRE observations, namely GOODS--South \citep{guo2013-a}, UDS \citep{galametz2013} and COSMOS \citep{nayyeri2017}. For the GOODS--North fied the CANDELS catalog was not yet finalized, and we used instead the \Ks-selected catalog of \cite{pannella2015}. Each of these fields is about $150\,{\rm arcsec}^2$ and they are evenly distributed on the sky to mitigate cosmic variance. We also used a catalog of the COSMOS $2\,\sq\degr$ field \citep{muzzin2013-a}, which has overall shallower data but covers a much larger area; this field provides important statistics for the rarest and brightest objects.

The ancillary photometry varies from one field to another, being a combination of both space- and ground-based imaging from various facilities. The UV to near-IR wavelength coverage typically goes from the $U$ band up the \spitzer IRAC $8\,\um$, including at least the \hst bands F606W, F814W, F125W, and F160W in CANDELS, and a deep $K$ (or $K_{\rm s}$) band. All these images are among the deepest available views of the sky. These catalogs therefore cover most of the important galaxy spectral features across a wide range of redshifts, even for intrinsically faint objects.

We augmented these catalogs with mid-IR photometry from \spitzer MIPS and far-IR photometry from \herschel PACS and SPIRE taken as part of the GOODS--\herschel \citep{elbaz2011}, CANDELS--\herschel programs (PI: M.~Dickinson), PEP \citep{lutz2011} and HerMES \citep{oliver2010}.

Photometric redshifts and stellar masses were computed following \cite{pannella2015} using EAzY \citep{brammer2008}; for COSMOS $2\,\sq\degr$ we used the redshifts from \cite{muzzin2013-a} which were computed the same way. For all catalogs, stellar masses were then computed using FAST \citep{kriek2009} by fixing the redshift to the best-fit photo-$z$ and fitting the observed photometry up to the IRAC $4.5\,\um$ band\footnote{The last two IRAC channels, at $5.8$ and $8\,\um$, were not used to derive the stellar mass for two reasons. First, at low redshift these bands are contaminated by the dust emission and AGNs, which cannot be taken into account by FAST. Second, while the corresponding images are reasonably deep in the two GOODS fields, the observations in UDS and COSMOS are substantially shallower. Excluding these bands from the fit therefore prevents introducing field-to-field systematics.} using the \cite{bruzual2003} stellar population synthesis model, assuming a \cite{salpeter1955} IMF, a \cite{calzetti2000} extinction law and a delayed exponentially-declining star formation history.

Galaxies with an uncertain photometric redshift (redshift \texttt{odds} less than $0.8$) or bad SED fitting (reduced $\chi^2$ larger than $10$) were excluded from our sample. The resulting sample is the one we used for stacking the \herschel images in \citetalias{schreiber2015}. In this previous work, we estimated the evolution of the stellar mass completeness (at the $90\%$ level) of these catalogs at all redshifts, and found that all the stacked samples with significant signal were complete in mass. For example, at $z=1$ the completeness is as low as $5\times10^8\,\msun$.

We estimated $\sfr$s by summing the emerging UV light and the dust obscured component observed in the mid- to far-IR, following \citep{daddi2007-a} and \citep{kennicutt1998-a} to convert the observed luminosities into star formation rates:
\begin{equation}
    \sfr = 2.17\times10^{-10}\,\luv\,[\lsun] + 1.72\times10^{-10}\,\lir\,[\lsun]\,.
\end{equation}
UV luminosities ($1500\,\text{\AA}$) were computed from the best-fit photo-$z$ template from EAzY, while IR luminosities were computed from the best-fit dust SED obtained with our new library (see \rsec{SEC:irsed_recipe}). We applied this method to both the stacked samples and to individual galaxies with mid- or far-IR detections. For individual galaxies, we did not attempt to measure the $\sfr$s of the rest of the sample without IR data since estimates based on the UV light alone are less reliable \citep{goldader2002,buat2005,elbaz2007,rodighiero2011,rodighiero2014}. When working with individual galaxies (as opposed to stacking), we therefore only considered the sub-sample of IR-detected galaxies, which implicitly corresponds to a $\sfr$ threshold at each redshift \citep[see][]{elbaz2011}. The resulting selection biases are discussed in \rsec{SEC:selection_effects}.

Lastly, the rest-frame $U$, $V$ and $J$ magnitudes were computed for each galaxy using EAzY, by integrating the best-fit galaxy template from the photo-$z$ estimation. These colors were used, following \cite{williams2009}, to separate galaxies that are ``quiescent'' from those that are ``star-forming''. We used the same selection criteria as those described in \citetalias{schreiber2015}, that is, a galaxy was deemed star-forming if its colors satisfy
\begin{equation}
    UVJ_{\rm SF} = \left\{\begin{array}{rcl}
        U - V &<& 1.3\,\text{, or} \\
        V - J &>& 1.6\,\text{, or} \\
        U - V &<& 0.88\times(V - J) + 0.49\,.
    \end{array}\right.\label{EQ:uvj}
\end{equation}
otherwise the galaxy was considered as quiescent as was thus excluded from the present study. As shown in \citetalias{schreiber2015}, only a very small fraction of the IR-detected galaxies are classified as \uvj quiescent, therefore this selection is only important for stacking.

\subsection{ALESS}

To improve the statistics on the dust temperature of individual galaxies at $z>1$, we complemented this sample with the $99$ galaxies observed by ALMA in the ALESS program \citep{hodge2013}. ALESS is a targeted program aiming to deblend the $870\,\um$ emission of sources detected in the single-dish LABOCA image of the ECDFS, which covers about $0.3\,\deg^2$ centered on the CANDELS GOODS--South field. The high resolution of the ALMA imaging allows a precise localization of the sub-millimeter source and avoids flux boosting from blended neighbors.

We used the photometric redshifts and stellar masses determined in \cite{dacunha2015} using Magphys \citep{dacunha2008}. We used the $24\,\um$ and PACS photometry from PEP, the SPIRE photometry as measured in \cite{swinbank2014} using the ALMA detections to improve the deblending, and the ALMA photometry from \cite{hodge2013}. The galaxies were then treated the same way as those from CANDELS, and used only in \rsec{SEC:irsed_indiv}.

\subsection{HRS}

To complement our sample toward the local Universe, we used the \herschel Reference Survey (HRS; \citealt{boselli2010}). This is a volume-limited survey targeting a few hundred galaxies in and out of the Virgo cluster, obtaining in particular \herschel PACS and SPIRE photometry for a full sampling of their dust SEDs. All the galaxies in the HRS also have UV-to-NIR coverage to determine stellar masses and colors. Of this sample, we only considered the \uvj star-forming galaxies which do not belong to the Virgo cluster, to avoid systematic effects caused by this peculiar environment, for a total of 131 galaxies with a minimum stellar mass of $10^{9}\,\msun$. This same sample was studied in \cite{schreiber2016}; further informations can be found there.

We used the photometry from \cite{ciesla2012} and modeled the HRS galaxies using CIGALE \citep{noll2009,roehlly2014}, which fits simultaneously the stellar and dust emission. Using CIGALE proved necessary because the contribution of the stellar continuum to the $8\,\um$ luminosity (or more generally to the PAH domain in the mid-IR) can be non-negligible at $z=0$, owing to the overall lower star-formation activity. We performed the fits using same SED libraries as for the CANDELS sample, that is, the dust SEDs introduced in this paper and the \citealt{bruzual2003} templates with a delayed SFH. Our dust SEDs are made available to all CIGALE users in the official package.

\section{A new far infrared template library \label{SEC:irsed}}

\subsection{Description of the model}

\begin{figure}
    \centering
    \includegraphics[width=9cm]{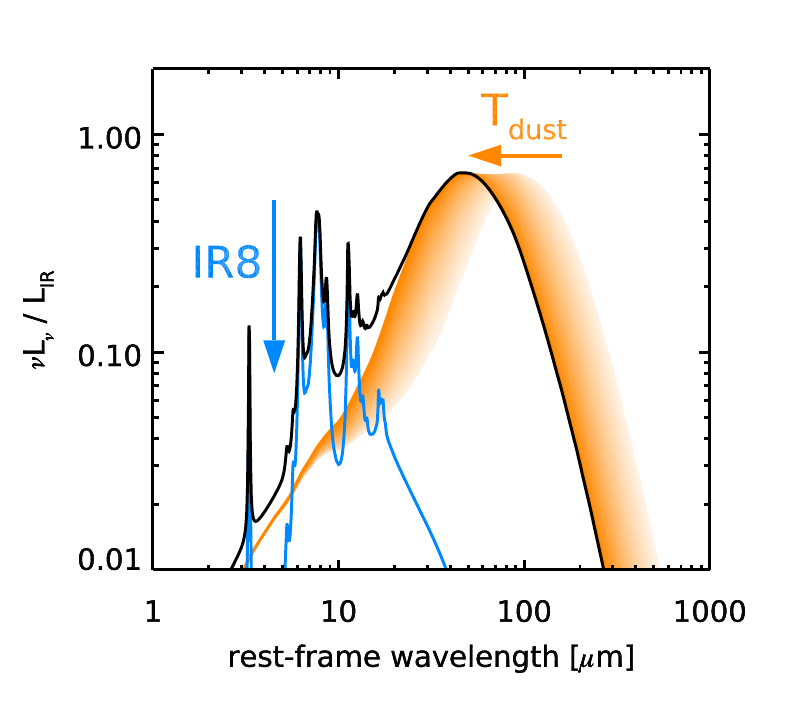}
    \caption{Cartoon picture illustrating the two effective parameters impacting the shape of our FIR SEDs. The total SED, normalized to unit $\lir$, is shown with a black solid line, while the dust continuum and PAH components are shown with solid orange and blue lines, respectively. We also show how the shape of the SED varies with dust temperature $\tdust$ by displaying several templates of different $\tdust$ in orange lines of varying intensities. The orange and blue arrows illustrate how the SED is modified by increasing $\tdust$ and $\ireight$, respectively.}
    \label{FIG:irseds}
\end{figure}

Since it was published, the \citetalias{chary2001} library has been used routinely to derive infrared luminosities, and therefore star formation rates, for large samples of galaxies at various redshifts. In \citetalias{schreiber2015}, we found that, in spite of the relatively small number of different SEDs it contains, it is able to fit well our stacked FIR photometry (rest-frame $30$-to-$500\,\um$) at all $z=0.5$ to $z=4$. However, once properly adjusted to the observed FIR data, the behavior of these SEDs, calibrated on local starbursts, may not adequately describe the observed MIR photometry. Indeed, the \citetalias{chary2001} library enforces a unique relation between $\tdust$, $\ireight$ and $\lir$ which was calibrated from Local Universe galaxies. The relevance of this assumption for distant galaxies was unknown at the time, and in fact it was shown to break when applied to $24\,\um$-detected $z=2$ galaxies \citep{papovich2007,daddi2007-a}. It was later understood that this was caused by an evolving $\lir$--$\ireight$ relation, and another more universal calibration was proposed where the $\ireight$ varies as a function of the distance to the main sequence \citep{elbaz2011,nordon2012,rujopakarn2013} rather than with the absolute luminosity. Similar conclusions have been drawn for the dust temperature \citep{elbaz2011,magnelli2014}.

To build a more up to date and versatile library, we started by making each of these observables independent of one another, that, we created a set of templates that allow us to vary $\tdust$, $\ireight$ and $\lir$ simultaneously. To do so, we made the arbitrary choice of separating the IR emission into two components: on the one hand, the dust continuum of varying $\tdust$ created by big and small grains, and on the other hand the MIR features emitted by PAH molecules (see \rfig{FIG:irseds}):
\begin{equation}
S_{\nu} = \mdust^{\rm cont}\,\bar{S}_{\nu}^{\rm cont} + \mdust^{\rm PAH}\,\bar{S}_{\nu}^{\rm PAH}\,. \label{EQ:sed_comp}
\end{equation}
$\mdust^{\rm PAH}$ and $\mdust^{\rm cont}$ are defined as the mass of dust grain found in the form of PAH molecules and silicate+carbonated grains, respectively, while $\bar{S}_{\nu}^{\rm PAH}$ and $\bar{S}_{\nu}^{\rm cont}$ are the spectra of each grain population normalized to unit dust mass (these are described in the next sections).

This decomposition implies that PAHs are almost exclusively responsible for the observed diversity in $\ireight$. Indeed, decomposing the MIR emission into multiple components (PAHs, very small grains, and AGN torus) is a degenerate problem when only broadband photometry is used, and a choice needs to be made. Since our objective is mostly to study galaxies and not AGNs, we neglect the presence of AGNs torus emission, and assume a fixed fraction of small vs.~big grains (see below). It is certainly possible to use our library in combination with an AGN template if sufficient MIR data is available, but this goes out of the scope of this work.

To create our templates, we used the amorphous carbon dust model of \cite{galliano2011} (hereafter \citetalias{galliano2011}). This model can output the mid- to far-IR spectrum emitted by a dust cloud of mass $1\,\msun$ under the influence of a uniform radiation field of integrated intensity $U$ (taken here in units of the \citealt{mathis1983} interstellar radiation field in the solar neighborhood, $U_\sun$). In the following, we call each spectrum generated by this model an ``elementary'' \citetalias{galliano2011} template. The next two sections describe how our library is built from these templates, as well as the underlying assumptions.

\subsubsection{Dust continuum}

In the \citetalias{galliano2011} model, the dust continuum elementary templates are produced by a combination of silicate and amorphous carbon grains of varying sizes, split into ``big'' (thermalized) and ``small'' (transiently heated) grains. The size distributions of these grains were taken from \cite{zubko2004} and were assumed to be universal. Here we assumed the Milky Way (MW) mass-fraction of carbonated vs.~silicate grains as derived by Zubko et al.~for big grains, but fixed the mass-fraction of small silicate grains to zero \citep[as in][]{compiegne2011} instead of $11\%$. While this is compatible with observational constraints \citep{li2001}, our motivation for this choice was purely empirical: reducing the emission of small grains in the mid-IR increased the range of $\ireight$ values that our model can reach.

The only remaining free parameter is the radiation intensity $U$. This parameter controls the energy of the radiation field to which each grain is exposed. For grains big enough to be thermalized, an increase of this energy implies an enhanced grain temperature (this is quantified later in \rsec{SEC:relations}), and affects the shape of their FIR spectrum. For smaller grains which are not thermalized, only the overall normalization of the spectrum is modified.

\subsubsection{PAH emission}

To produce the associated PAH emission, we assumed that these molecules are subject to the same $U$ as the other dust grains, although in this case this choice has very little consequence since, as for small grains, the PAH molecules are not thermalized and therefore the effect of increasing $U$ is essentially only to increase the normalization of the PAH SED at fixed dust mass. This will affect the absolute values of the PAH masses, in which we have limited interest here. The only free parameter in the \citetalias{galliano2011} model regarding the PAH composition is the fraction of neutral vs.~ionized molecules, which we chose here to be the MW value of $50\%$ \citep{zubko2004}. This parameter mostly influences the relative strengths of the $8$ vs.~$12\,\um$ PAH features, and we found that the MW value provided indeed a good match to the stacked \spitzer IRS spectra of $z=2$ ULIRGs (see \rsec{SEC:irs_spec}), as well as to our stacked $S_{24}/S_{\!16}$ broadband flux ratio at $z=1$ (see \rsec{SEC:irsed_stack}).

\subsubsection{Radiation field distribution}

The elementary templates introduced above are not well suited to describe an entire galaxy, since the hypothesis of a uniform $U$ usually does not hold in such kind of extended systems. Instead, a ``composite'' template must be built by adding together the emission of different dusty regions, heated by different radiation intensities. As in \cite{dale2001}, \citetalias{galliano2011} assumed that the distribution of $U$ in a composite system follows a power law in $\dd \mdust/ \dd U \sim U^{-\alpha}$, where $\dd \mdust$ is the mass of dust associated to the elementary region illuminated with an intensity $[U, U+\dd U]$. This distribution is then integrated from $U=U_{\rm min}$ to $U=U_{\rm max}$ to form the final template. Contrary to \citetalias{draine2007}, they did not assume the presence of an additional component linked to photo-dissociation regions since it was shown not to provide significant improvement to the fits (as demonstrated in the Appendix of \citetalias{galliano2011}).

The parameter $\alpha$ is the slope of the mass distribution of $U$ within a galaxy. This parameter affects the composite spectrum in a non-trivial way: large and small values of $\alpha$ will accumulate most of the dust mass close to $U_{\rm min}$ and $U_{\rm max}$, respectively ($\alpha=1$ and $2$ give a uniform weighting in mass and luminosity, respectively). To avoid this complexity, we fixed $\alpha=2.6$ for all our templates. This value was chosen to reproduce the width of the \citetalias{chary2001} SEDs between $15$ and $500\,\um$, since these templates are known to provide a good description of the FIR emission of distant galaxies \citep{elbaz2010}. We also checked a posteriori that this value provided a satisfactory fit to our stacked FIR photometry (see \rsec{SEC:irsed_stack}).

With our adopted $U$ distribution, the final SED is relatively insensitive to the precise choice of $U_{\rm max}$, provided $U_{\rm max} \gg U_{\rm min}$ \citep{draine2007-a}. Hence the only remaining parameter that allows us to tune the SED shape is $U_{\rm min}$ or, equivalently, the mass-weighted intensity $\mean{U}$. In particular, we note in \rsec{SEC:relations} that $\mean{U}$ is related to the average dust temperature through a simple power law, which is one of the parameter our library aims to describe. Therefore, we generated a logarithmic grid of $U_{\min}$ ranging from $0.1$ to $5\,000\,U_\sun$ with $250$ samples, and took $U_{\rm max} = 10^6\,U_\sun$ \citep{draine2007-a}. This allows our library to describe dust temperatures ranging from about $15$ to $100\,\kelvin$ with a roughly constant step of $0.3\,\kelvin$. The resulting SEDs can be obtained on-line (a link is provided on the first page).

\subsubsection{Amorphous carbon or graphite? \label{SEC:amorphous}}

Compared to more standard dust models (e.g., \citetalias{draine2007}), the one we used here assumes that carbonated grains are found exclusively in the form of amorphous carbon grains, rather than graphites. While this has no visible impact on the shape of the generated spectra, it systematically lowers the value of the measured dust masses by a factor of about $2.0$ compared to graphite dust, owing to the different emissivities of these grain species. This was in fact the motivation for using amorphous carbon in \citetalias{galliano2011}: lowering the measured dust masses eases the tension between the observed dust-to-gas ratio and stellar abundances in the Large Magellanic Cloud (LMC). This conclusion was not only reached in the LMC, which has a particularly low metallicity, but also in more normal galaxies including the MW \citep[e.g.,][]{compiegne2011,jones2013,fanciullo2015,planckcollaboration2016}. As stressed in \citetalias{galliano2011}, purely amorphous carbon is just one possibility to achieve higher emissivities. We therefore do not give much credit to carbon dust being truly amorphous, but since this type of grains does describe the content of the ISM in a more consistent way, we chose to favor it instead of graphite. The impact of this choice on gas-to-dust ratios is discussed in \rsec{SEC:gas}.

\subsection{Basic usage of the library and useful relations \label{SEC:relations}}

In this section we provide a set of simple relations to relate the internal parameters of the library, namely $\mean{U}$, $\mdust^{\rm cont}$, and $\mdust^{\rm PAH}$, to more commonly used observables such as the total infared luminosity, the dust mass, and the $\ireight$. We then provide instructions for the most basic usage of this template library.

\subsubsection{Dust temperature}

The dust temperature of our model SEDs was computed by applying Wien's law to each elementary \citetalias{galliano2011} template for the dust continuum:
\begin{equation}
\tdust [\kelvin] = 2.897\times10^{3}/(\lambda_{\rm max} [\um])\,,
\end{equation}
determining $\lambda_{\rm max}$ as the wavelength corresponding to the peak of $\lambda^\beta L_{\nu}$ (the term $\lambda^\beta$ takes into account the effective emissivity of the templates, $\beta \simeq 1.5$). We then weighted each value by the dust mass associated to the corresponding template ($\dd \mdust$), therefore producing a mass-weighted average. We found that the following relation links together $\tdust$ and the radiation field intensity:
\begin{equation}
\frac{\mean{U}}{U_\sun} = \left(\frac{\tdust}{18.2\kelvin}\right)^{5.57}\,. \label{EQ:tdust_umean}
\end{equation}
As stated earlier, our library covers $\tdust$ values ranging from $15$ to $100\,\kelvin$. We also applied Wien's law (as above) to the peak of the final dust template, to obtain a light-weighted average $\tdust^{\rm light}$. In practice, we find the difference between the two to be simply a constant factor, with
\begin{equation}
\tdust = 0.91\times\tdust^{\rm light}\,. \label{EQ:tdust_weight}
\end{equation}
$\tdust^{\rm light}$ is less stable because the summed dust template is broader, making it harder to accurately locate the position of the peak. Its physical meaning is also less clear, since our templates do not have a single temperature, but it has the advantage of being less model-dependent and is essentially the temperature one would measure by using a modified blackbody model with a single temperature and an emissivity of $\beta=1.5$. We therefore provide tabulated values for both temperatures in the library, and in the following, unless otherwise stated, we will refer to the dust temperature as the mass-weighted value. Likewise, using \req{EQ:tdust_umean} we mapped a dust temperature to each value of $\mean{U}$ and will refer to the two quantity interchangably.

\subsubsection{Total infrared and $8\,\um$ luminosities}

Each template and component in our library, both for the continuum and PAH emission, is associated with a value of $\lir$ ($8$-to-$1000\,\um$) and $\leight$ (integrated with the response of the \spitzer IRAC channel 4, i.e., $6.4$-to-$9.3\,\um$). Both luminosities were computed as the integral of $L_\lambda\,\dd\lambda$ within their respective wavelength interval, and are given in units of total solar luminosity ($\lsun=3.839\times10^{26}\,{\rm W}$). The luminosities are proportional to the mass of dust, and depend linearly on $\mean{U}$:
\begin{align}
\lir &=  \lir^{\rm cont} + \lir^{\rm PAH}, \\
\lir^{\rm cont} [\lsun] &= 191 \, (\mdust^{\rm cont}/\msun) \, (\mean{U}\!/U_\sun), \label{EQ:lir_mdust} \\
\lir^{\rm PAH} [\lsun] &= 325 \, (\mdust^{\rm PAH}/\msun) \, (\mean{U}\!/U_\sun), \\
\leight &=  \leight^{\rm cont} + \leight^{\rm PAH}, \\
\leight^{\rm cont} [\lsun] &= 7.05 \, (\mdust^{\rm cont}/\msun) \, (\mean{U}\!/U_\sun), \\
\leight^{\rm PAH} [\lsun] &= 755 \, (\mdust^{\rm PAH}/\msun) \, (\mean{U}\!/U_\sun).
\end{align}
Defining the mass fraction of PAHs as $\fpah \equiv \mdust^{\rm PAH}/\mdust$, we have
\begin{align}
\ireight \equiv \frac{\lir}{\leight} &= \frac{\lir^{\rm cont}\,(1-\fpah) + \lir^{\rm PAH}\,\fpah}{\leight^{\rm cont}\,(1-\fpah) + \leight^{\rm PAH}\,\fpah}\,, \nonumber \\
\frac{1}{\fpah} &= 1 - \frac{\lir^{\rm PAH} - \leight^{\rm PAH}\,\ireight}{\lir^{\rm cont} - \leight^{\rm cont}\,\ireight}\,. \label{EQ:fpah_ir8_th}
\end{align}
Using the above approximate equations, this becomes:
\begin{align}
\ireight = \frac{191 + 134\,\fpah}{7.05 + 748\,\fpah}\, \quad \text{and} \quad
\fpah = \frac{191 - 7.05\,\ireight}{-134 + 748\,\ireight}\,. \label{EQ:fpah_ir8}
\end{align}
By varying the relative contribution of PAHs to the total dust mass, our library can reach $\ireight$ values in the range $0.48$ to $27.7$, which covers the vast majority of the observed parameter space \citep{elbaz2011}.

\subsubsection{Basic usage}

Our library is composed of multiple templates, each corresponding to a given dust temperature $\tdust$. As illustrated in \rfig{FIG:irseds}, each template is composed of two components: dust continuum on the one hand, and PAH emission on the other hand. The amplitude of each component is internally dictated by the corresponding mass of dust grains, $\mdust^{\rm cont}$ and $\mdust^{\rm PAH}$, respectively (\req{EQ:sed_comp}), and both can be freely adjusted to match the observed data, effectively varying the dust mass (or $\lir$) and the $\ireight$.

To perform the fit, both PAH and continuum components must be redshifted to the assumed redshift of the source, and the expected flux in each observed passband is computed by integrating the redshifted template multiplied with the corresponding filter response curve. At this stage, the expected fluxes can be fit to the observed ones through a linear solver, varying simultaneously $\mdust^{\rm cont}$ and $\mdust^{\rm PAH}$. By performing such a fit for each value of $\tdust$ in the library and picking the smallest $\chi^2$, one can find the optimal model (in the $\chi^2$ sense) corresponding to the provided photometry. This is the simplest way to use the library, and it will work in most cases if enough photometry is available and if the redshift is precisely known. In other cases, a more careful approach should be used, and we describe it later in \rsec{SEC:irsed_recipe}.

The immediate products of this fit are the dust masses, $\mdust^{\rm cont}$ and $\mdust^{\rm PAH}$ (expressed in solar mass), and the dust temperature $\tdust$. The total dust mass can be obtained by summing the two components, $\mdust = \mdust^{\rm cont} + \mdust^{\rm PAH}$, while $\lir$ and $\leight$ are tabulated (per unit solar mass of dust) for each $\tdust$ value in the library, or can be estimated using the above relations. In the next section, we check the accuracy of our PAH templates by fitting stacked MIR spectra from the \spitzer IRS spectrometer.

\subsection{Comparison against stacked MIR spectroscopy \label{SEC:irs_spec}}

\begin{figure}
    \centering
    \includegraphics[width=0.5\textwidth]{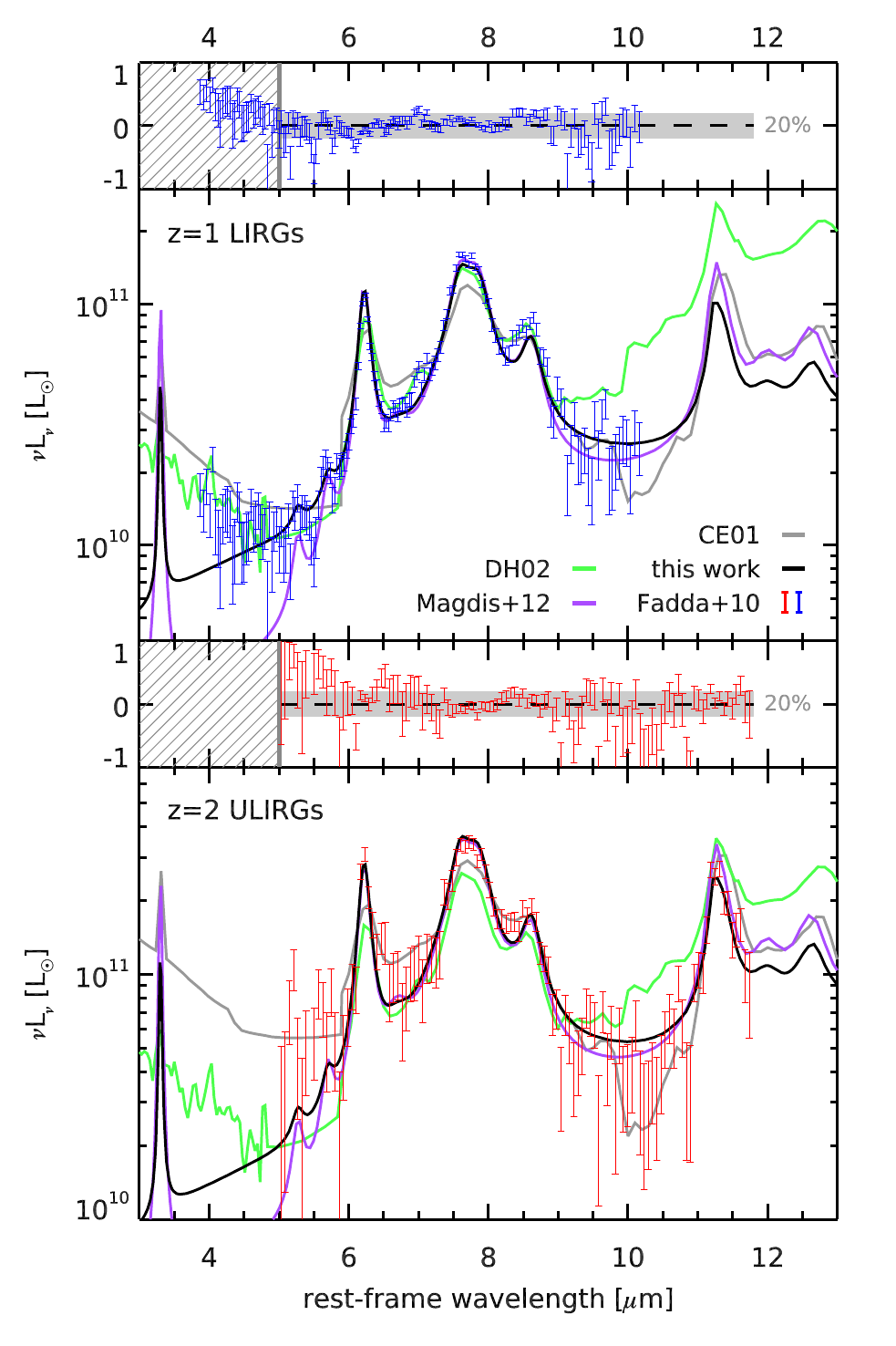}
    \caption{Comparison of the our templates (black solid line) against stacked \spitzer IRS spectra of $z=1$ LIRGs (blue, top) and $z=2$ ULIRGs (red, bottom) from \cite{fadda2010}. The relative residuals of the fits are shown above the plot for each sample; the region of perfect agreement is shown with a dashed line, surrounded by a $\pm20\%$ confidence interval. We also show the best fit using other models: \citetalias{chary2001} (gray), \citetalias{dale2002} (green), and \cite{magdis2012} (purple). In all cases the fit only uses observations at $\lambda > 5\,\um$, since shorter wavelength can be contaminated by the stellar continuum, as illustrated with the hashed region in the residual plots.}
    \label{FIG:irs_spec}
\end{figure}

During the cryogenic phase of the \spitzer mission, the IRS spectrometer could observe in the MIR from $5.3$ to $38\,\um$ at low ($R=90$) and medium ($R=600$) spectral resolutions. It has therefore provided valuable measurements of the PAH emission, both in local and distant galaxies. In particular, \cite{fadda2010} have observed a sample of $z=1$ LIRGs and $z=2$ ULIRGs in the GOODS--South field (plus a few in the wider ECDFS). The galaxies in these two samples have been selected based on their redshift and \spitzer MIPS $24\,\um$ flux ($0.2$--$0.5\,\mJy$), therefore their measured properties cannot be straightforwardly compared to the mass-complete stacks that we will analyze in \rsec{SEC:irsed_stack}. Because of the $24\,\um$ flux limit, the sample of Fadda et al.~is biased toward starburst galaxies located above the main sequence. With this caveat in mind, this dataset can still be used as a consistency check for our SED library. Indeed, although these samples may be biased, our library must be able to at least broadly reproduce their PAH spectra.

We therefore applied the fitting method described in the previous section to the stacked spectra of both $z=1$ and $z=2$ samples. Since the dust temperature cannot be constrained from IRS spectroscopy alone, we fixed it to its redshift-average value of $\tdust=28$ and $33\,\kelvin$, respectively (see \rsec{SEC:irsed_stack}), although this choice had no impact on the quality of the fit. We also did not attempt to fit rest-frame wavelengths $\lambda<5\,\um$ to avoid contamination from stellar continuum. The result is shown in \rfig{FIG:irs_spec}, together with fits from a number of other libraries from the literature. It can be seen from this figure that we are able to fit these spectra with good accuracy (typically better than $20\%$) and obtain a significantly improved match compared to the \citetalias{chary2001} or \citetalias{dale2002} libraries. The \cite{magdis2012} library (which uses the \citetalias{draine2007} models) yields a similarly good fit as ours, with only subtle differences. This should not come as a surprise, since the physical properties of the PAHs in both the Magdis et al.~library and ours are adapted from \citetalias{draine2007}.

\section{Observed dust temperatures and IR8 \label{SEC:calib}}

\subsection{Average values from stacked photometry \label{SEC:irsed_stack}}

\begin{figure*}
    \centering
    \includegraphics[width=0.8\textwidth]{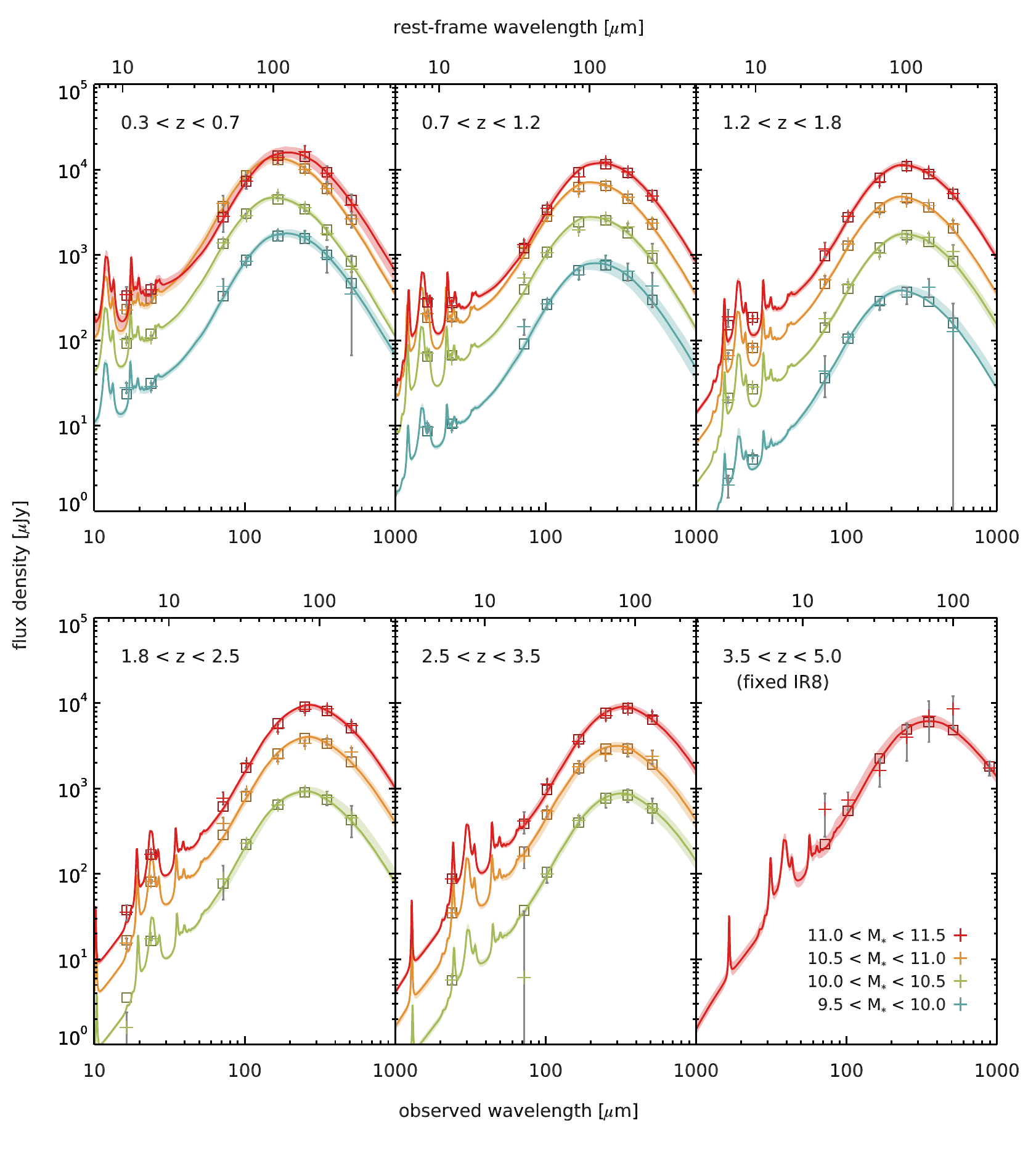}
    \caption{\spitzer and \herschel stacks of \citetalias{schreiber2015} (crosses) of main sequence galaxies at different redshifts (from left to right) and for different stellar masses (colors, see legend). These also include new stacks of the \spitzer $16\,\um$ and \herschel $70\,\um$ images in the GOODS fields, as well as ALMA $890\,\um$ from \cite{schreiber2017-a} for the $z=4$ bin. We overplot the best fit template from our library with colored solid lines. Open boxes in the background show the modeled broadband flux from the best-fit template, to illustrate any offset with the observations. The colored regions in the background display the scatter of the best-fit model SED among all the bootstrap realizations.}
    \label{FIG:irstacks}
\end{figure*}

\subsubsection{Description of the stacked data}

We now proceed to apply our new library to model the stacked \spitzer and \herschel photometry in the CANDELS fields. These stacks were presented in detail in \citetalias{schreiber2015}, and we briefly recall here the essential information. We selected all galaxies in the GOODS--North, GOODS--South, UDS and COSMOS CANDELS fields that are brighter than $H=26$ and \uvj star-forming (\req{EQ:uvj}). These galaxies were then binned according to their redshift and stellar mass, for a total of $24$ bins (six redshift bins from $z=0.3$ to $5$, and four mass bins from $\mstar=3\times10^9$ to $3\times10^{11}\,\msun$). The number of stacked galaxies in each bin is given in Figure 4 of \citetalias{schreiber2015}. The smallest number of stacked galaxy were $53$ and $28$, in the highest mass bins at $z\sim 0.5$ and $z\sim 4$, respectively. Otherwise, all the bins had at least $100$ galaxies (median of $628$).

In the bins where we estimated our stellar-mass completeness to be above $90\%$, we stacked the \spitzer and \herschel images at the positions of all the galaxies in the bin. The fluxes in the resulting stacked images were measured through point spread function (PSF) fitting with a free background. The measured fluxes were corrected for clustering {\it a posteriori} using an empirical recipe calibrated on simulated images, and the method is described fully in the appendix of \citetalias{schreiber2015}. Briefly, we simulated the \herschel maps using the positions of the real galaxies of CANDELS and the prescriptions described in \cite{schreiber2017-b} to predict (statistically) their $\lir$ and FIR fluxes. The resulting images have the same statistical properties (pixel distribution and number counts) as the real images. We then applied the same stacking method and compared the measured fluxes to the true flux averages in the stacked sample. We found that the flux boosting caused by clustering is roughly constant in a given band, hence we assumed a fixed correction in each \herschel band throughout, at all redshifts and masses (see also \citealt{bethermin2015-a}). The largest correction was a reduction of the fluxes by $25\%$ for the SPIRE $500\,\um$ band.

The uncertainty on each flux measurement was finally computed by taking the maximum value of two independent estimates: first by doing bootstrapping, that is, repeatedly removing half of the sample from the stack, and second from the RMS of the residual image after subtraction of the best-fit PSF model.

In \citetalias{schreiber2015}, we only stacked the \spitzer MIPS $24\,\um$ band as well as the \herschel bands redward of $100\,\um$, which are the only bands covered in all the four CANDELS fields. For the present work, we extended these stacks to also include the \spitzer IRS $16\,\um$ imaging \citep[acquired in the GOODS fields only;][]{teplitz2011} as well as the \herschel PACS $70\,\um$ (acquired in the GOODS--South field only). Because these images only cover some of the CANDELS fields, the stacked fluxes can be affected by field-to-field variations. To correct for this effect, we first computed the $S_{16}/S_{24}$ flux ratio observed when stacking only the two GOODS fields. We then multiplied the stacked $24\,\um$ flux obtained with the four fields by this ratio to estimate the corresponding $16\,\um$ flux. The same procedure is used for the $70\,\um$ flux, based on the $S_{70}/S_{100}$ flux ratio observed in GOODS--South. Lastly, to better constrain the dust temperature for our $z=4$ bin we added the average ALMA $890\,\um$ of our galaxies as measured in \cite{schreiber2017-a} (in all fields but GOODS--North).

\subsubsection{Fitting of the stacked SEDs}

Using the fitting method described in \rsec{SEC:relations}, we used our new library to model the stacked fluxes from \citetalias{schreiber2015} (avoiding again $\lambda<5\,\um$). We used the mean redshift of the stacked sample to shift the SED in the observed frame. Uncertainties on the redshifts are of the order of $5\%$ and are thus much smaller than the bins (which have a constant width of $25\%$), we therefore ignored them, but we did broaden the templates by the width of the redshift bins. To derive accurate error estimates on the fitting parameters, we bootstrapped the stacked sample and applied the fitting procedure on each bootstrapped SED. This allows a better treatment of correlated noise (flux fluctuations affecting multiple bands simultaneously due, e.g., to contamination from neighboring sources on the map). The resulting models are compared to the measured photometry in \rfig{FIG:irstacks}.

The first thing to note is the excellent agreement between our templates and the photometry. With only three free parameters, no clear tension is observed, reinforcing the idea of a universal SED. Compared to our previous fits with the \citetalias{chary2001} library, we found very similar values of $\lir$. The most extreme difference arose in the lowest redshift bin ($0.3 < z < 0.7$) where we obtained values that are systematically $0.1\,\dex$ lower with the new library. This difference is caused by a peculiar feature of the previously adopted best-fit \citetalias{chary2001} template. This particular SED (ID $40$) shows an enhanced flux around the rest-frame $30\,\um$ compared to our library. Without any data to constrain this feature, we cannot say whether it is real or not, although we tend to favor the result of our new SED library which has a consistent shape at all $\tdust$.

\subsubsection{Best fit parameters and systematic bias corrections \label{SEC:stack_correct}}

One major issue when interpreting stacked photometry is that the stacked SED is the flux-weighted average in each band, which is not necessarily equal to the SED of the average galaxy of the sample. This is particularly true if the brightest galaxies have a different SED than the average galaxy, and indeed such situation is expected in our case: starburst galaxies, which have the highest $\sfr$s in a given bin of mass, have an increased temperature and $\ireight$. Therefore our stacked SED will be biased towards higher temperatures and higher $\ireight$ compared to their true average, and these biases need to be accounted for before going further.

To quantify these biases, we used the Empirical Galaxy Generator (EGG; \citealt{schreiber2017-b}). Using a set of empirical prescriptions, this tool can generate mock galaxy catalogs matching exactly the observed stellar mass functions at $0<z<6$ and the galaxy main sequence with its scatter and starburst population (\citetalias{schreiber2015}). We also implemented the relations derived in the present paper for $\tdust$ and $\ireight$ and their dependence on $\rsb$, namely \reqs{EQ:tdust_ms} to \ref{EQ:ir8_sb}, adding the observed residual scatter (see \rsecs{SEC:indiv_tdust_result} and \ref{SEC:indiv_ir8_result}) as a random Gaussian perturbation to match the observed distribution of both quantities. Thanks to these prescriptions, the tool can produce realistic mock catalogs of galaxies with a full infrared SED, and we showed in \cite{schreiber2017-b} that this empirical model reproduces faithfully the observed number counts in all FIR bands.

Generating a large mock catalog of about a million galaxies with $\mstar > 3\times 10^{9}\,\msun$ at $0.3<z<5$, we simulated our stacked SEDs by computing the average flux in each band for each of the stellar mass and redshift bins displayed in \rfig{FIG:irstacks}. The size of the mock catalog was chosen so that each stacked bin contained at least $1\,000$ galaxies. No noise was added to the stacked fluxes, since we only looked for systematic biases. We then applied exactly the same fitting method to these synthetic stacks as used for the real stacks, and compared the resulting fit parameters to their true average. Since the recipes for $\tdust$ and $\ireight$ used in EGG depend to some extent on the bias correction we are now discussing, we proceeded iteratively: we derived a first estimate of \reqs{EQ:tdust_ms} to \ref{EQ:ir8_sb} without applying any bias correction, implemented these relations in EGG, and determined a first value of the corrections. We then applied these corrections to the observed values, re-evaluated the relations, updated EGG and the mock catalog, then determined the final corrections. The obtained values were not significantly different from the first estimates, so we only did two iterations of this procedure.

We found that the ``raw'' stacked values (before correction) were on average larger than their true average by $1.5\pm0.3\,\kelvin$ for $\tdust$, and by a factor of $10\pm1\%$ for $\ireight$. These are relatively small corrections which do not impact our conclusions, but we applied them nevertheless to our best fit values.

\begin{figure*}
    \centering
    \includegraphics[width=\textwidth]{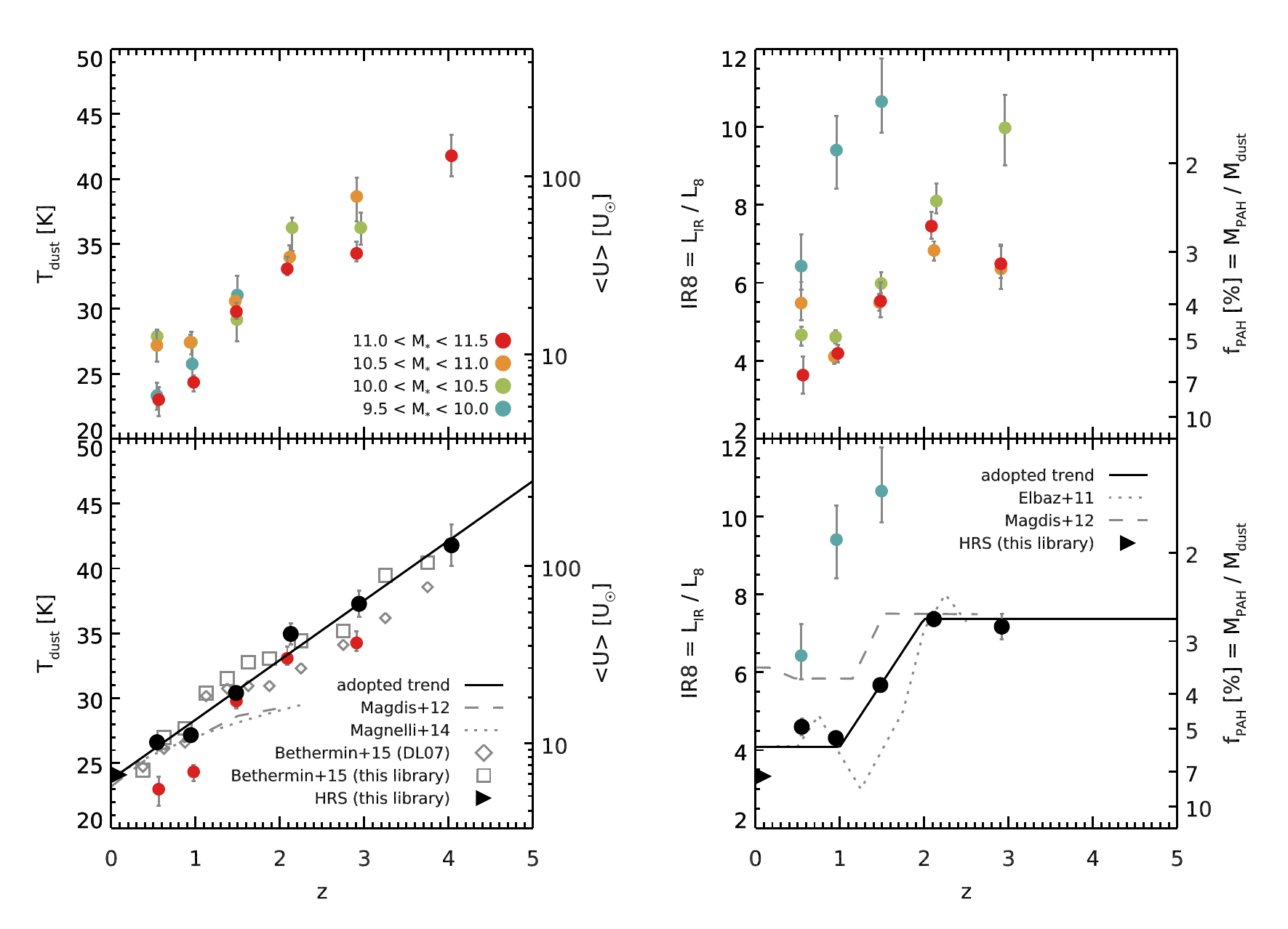}
    \caption{{\bf Left:} Evolution of the effective dust temperature $\tdust$ with redshift. The $\tdust$ estimated from each stacked SED at different stellar masses are shown on the top plot with circles of different colors (see legend). The bottom plot shows the weighted mean of all mass bins as black circles (excluding the most massive bin, shown in red, or the least massive bin, shown in blue). The average of the galaxies from the HRS \citep{schreiber2016,ciesla2014} is given with a solid triangle. The trend we adopt in this paper is shown with a solid black line. We also show the $\tdust$ evolution of \cite{magdis2012} and \cite{bethermin2015-a} (both converted from $\mean{U}$ to $\tdust$ using \req{EQ:tdust_umean}) as well as \cite{magnelli2014} (corrected from light-weighted to mass-weighted using \req{EQ:tdust_weight}). We also show the best fitting temperature to the \cite{bethermin2015-a} SEDs modeled with our own library as open squares. {\bf Right: } Evolution of the $\ireight$ ratio with redshift. The legends are the same as for the plots on the left, except that here we show the values obtained by \cite{magdis2012} (computed from their best-fit template SEDs) and \cite{elbaz2011}.}
    \label{FIG:tdust_fpah_stack}
\end{figure*}

\begin{table*}
\centering
\caption{Best fit dust parameters for the stacked SEDs. \label{TAB:stack_best_fits}}
\begin{tabular}{ccccccc}
\hline \\[-0.35cm]
$z$        & $\log_{10}\mstar$       & $\lir$               & $\mdust$             & $\tdust$             & $\fpah$             & $\ireight$   \\
           & $\log_{10}\msun$ & $10^{10}\lsun$       & $10^7\msun$          & $\kelvin$            & \%                  &              \\        \hline\hline \\[-0.35cm]
0.3 -- 0.7 & 9.5 -- 10.0  & $1.49^{+0.08}_{-0.09}$ & $0.136^{+0.034}_{-0.025}$ & $23.3^{+1.0}_{-1.1}$    & $3.08^{+0.40}_{-0.46}$ & $6.4^{+0.8}_{-0.6}$ \\[0.1cm]
           & 10.0 -- 10.5 & $4.61^{+0.20}_{-0.20}$ & $0.180^{+0.019}_{-0.017}$ & $27.9^{+0.5}_{-0.4}$    & $4.66^{+0.36}_{-0.28}$ & $4.66^{+0.22}_{-0.28}$ \\[0.1cm]
           & 10.5 -- 11.0 & $12.9^{+0.9}_{-0.9}$   & $0.57^{+0.10}_{-0.07}$    & $27.2^{+1.2}_{-1.2}$    & $3.8^{+0.5}_{-0.5}$    & $5.5^{+0.5}_{-0.4}$ \\[0.1cm]
           & 11.0 -- 11.5 & $14.9^{+1.5}_{-1.5}$   & $1.43^{+0.38}_{-0.26}$    & $23.0^{+1.0}_{-1.3}$    & $6.4^{+0.9}_{-0.9}$    & $3.6^{+0.5}_{-0.5}$ \\[0.1cm]
0.7 -- 1.2 & 9.5 -- 10.0  & $2.24^{+0.13}_{-0.18}$ & $0.130^{+0.064}_{-0.035}$ & $25.7^{+1.8}_{-1.7}$    & $1.76^{+0.35}_{-0.24}$ & $9.4^{+0.9}_{-1.0}$ \\[0.1cm]
           & 10.0 -- 10.5 & $8.48^{+0.23}_{-0.35}$ & $0.354^{+0.061}_{-0.031}$ & $27.4^{+0.8}_{-0.9}$    & $4.73^{+0.26}_{-0.25}$ & $4.61^{+0.16}_{-0.17}$ \\[0.1cm]
           & 10.5 -- 11.0 & $21.2^{+0.8}_{-0.7}$   & $0.88^{+0.08}_{-0.08}$    & $27.4^{+0.6}_{-0.4}$ & $5.47^{+0.33}_{-0.26}$ & $4.11^{+0.15}_{-0.19}$ \\[0.1cm]
           & 11.0 -- 11.5 & $35.2^{+1.8}_{-2.3}$   & $2.58^{+0.31}_{-0.31}$    & $24.3^{+0.5}_{-0.7}$    & $5.36^{+0.33}_{-0.31}$ & $4.19^{+0.22}_{-0.23}$ \\[0.1cm]
1.2 -- 1.8 & 9.5 -- 10.0  & $2.91^{+0.22}_{-0.14}$ & $0.063^{+0.010}_{-0.016}$ & $31.1^{+1.5}_{-0.3}$ & $1.40^{+0.23}_{-0.27}$ & $10.7^{+1.1}_{-0.8}$ \\[0.1cm]
           & 10.0 -- 10.5 & $13.0^{+0.6}_{-0.5}$   & $0.40^{+0.10}_{-0.06}$    & $29.2^{+0.3}_{-1.7}$ & $3.36^{+0.20}_{-0.21}$ & $5.98^{+0.30}_{-0.26}$ \\[0.1cm]
           & 10.5 -- 11.0 & $36.3^{+1.3}_{-1.5}$   & $0.85^{+0.07}_{-0.07}$    & $30.6^{+0.3}_{-0.5}$ & $3.75^{+0.19}_{-0.21}$ & $5.50^{+0.22}_{-0.21}$ \\[0.1cm]
           & 11.0 -- 11.5 & $84^{+4}_{-4}$         & $2.35^{+0.37}_{-0.24}$    & $29.8^{+0.7}_{-0.6}$    & $3.72^{+0.38}_{-0.38}$ & $5.5^{+0.5}_{-0.4}$ \\[0.1cm]
1.8 -- 2.5 & 10.0 -- 10.5 & $16.3^{+0.7}_{-0.8}$   & $0.131^{+0.069}_{-0.024}$ & $36.2^{+0.8}_{-1.8}$    & $2.13^{+0.13}_{-0.19}$ & $8.10^{+0.45}_{-0.32}$ \\[0.1cm]
           & 10.5 -- 11.0 & $65.4^{+2.3}_{-2.8}$   & $0.76^{+0.16}_{-0.13}$    & $34.0^{+0.9}_{-1.1}$    & $2.75^{+0.12}_{-0.14}$ & $6.83^{+0.24}_{-0.26}$ \\[0.1cm]
           & 11.0 -- 11.5 & $144^{+6}_{-7}$        & $2.04^{+0.28}_{-0.28}$    & $33.1^{+0.9}_{-0.5}$    & $2.44^{+0.15}_{-0.20}$ & $7.46^{+0.36}_{-0.32}$ \\[0.1cm]
2.5 -- 3.5 & 10.0 -- 10.5 & $26.5^{+2.2}_{-2.0}$   & $0.213^{+0.069}_{-0.033}$ & $36.2^{+1.2}_{-1.3}$    & $1.52^{+0.30}_{-0.21}$ & $10.0^{+0.9}_{-1.0}$ \\[0.1cm]
           & 10.5 -- 11.0 & $102^{+7}_{-10}$       & $0.58^{+0.14}_{-0.13}$    & $38.6^{+1.4}_{-1.9}$    & $2.98^{+0.36}_{-0.37}$ & $6.4^{+0.6}_{-0.5}$ \\[0.1cm]
           & 11.0 -- 11.5 & $257^{+17}_{-16}$      & $2.9^{+0.5}_{-0.5}$       & $34.3^{+0.9}_{-0.6}$    & $2.96^{+0.25}_{-0.25}$ & $6.49^{+0.46}_{-0.37}$ \\[0.1cm]
3.5 -- 5.0 & 11.0 -- 11.5 & $357^{+37}_{-61}$      & $1.41^{+0.26}_{-0.18}$    & $41.8^{+1.6}_{-1.6}$    & ---                    & --- \\[0.1cm]
\hline
\end{tabular}
\end{table*}

In \rfig{FIG:tdust_fpah_stack} (top left and top right) we show the best-fit values we obtain for $\tdust$ and $\ireight$ in all bins of mass and redshifts where they could be measured. These are also tabulated in \rtab{TAB:stack_best_fits}. To complement our stacks toward the local Universe, we also compute the average $\tdust$ and $\ireight$ for the \uvj star-forming galaxies of the HRS. The evolution with redshift and mass of $\tdust$ and $\ireight$ are quantified in the following sections.

\subsubsection{Evolution of the dust temperature}

In most cases, varying the stellar mass has no influence on the dust temperature. The most massive galaxies ($\mstar > 10^{11}\,\msun$) tend to have colder temperatures by $2.3\,\kelvin$ on average (see also \citealt{matsuki2017} who report a similar trend at $z=0$), but this is only significant at $z<1$, in the domains where the main sequence departs from a linear relation \citep[e.g.,][]{whitaker2015,schreiber2015}. This reduced temperature was already interpreted in our previous work as a sign that massive star-forming galaxies at low redshifts are in the process of shutting down star-formation, with a slowly declining efficiency \citep{schreiber2016}. Since this is a minor effect compared to the overall redshift evolution, and since it only affects the few most massive galaxies, we do not discuss it further here.

Averaging the dust temperatures of the mass bins unaffected by the above effect, we recovered a previously reported trend for the dust temperature to increase with redshift \citep[e.g.,][]{magdis2012,magnelli2013,bethermin2015-a}. Contrary to what \cite{magdis2012} claimed, we observed a continuous rise of the temperature up to $z=4$, confirming the results of \cite{bethermin2015-a}. Fitting the evolution of $\tdust$ with redshift as an empirical power law, we obtained
\begin{equation}
\tdust^{\rm MS} [\kelvin] = (32.9 \pm 2.4) + (4.60 \pm 0.35) \times (z - 2)\,.\label{EQ:tdust_ms}
\end{equation}
This relation is displayed in \rfig{FIG:tdust_fpah_stack} (bottom left), where we compare it to results from the literature. In particular \cite{magnelli2014} found $\tdust = 26.5 \times (1+z)^{0.18}$. The normalization of this relation is higher than the one we report here, but the evolution with redshift is milder. This higher normalization is linked to the fact that \cite{magnelli2014} measured $\tdust$ by fitting modified blackbodies, making their dust temperatures light-weighted. Correcting for this difference using \req{EQ:tdust_weight} (as was done in \rfig{FIG:tdust_fpah_stack}), their $z=0$ value is fully consistent with ours, but their measurement at $z=2$ falls short of ours by $5\kelvin$. The same is true for the stacks of \cite{magdis2012}. This could be caused by the absence of clustering correction in these two studies, since it affects preferentially the long wavelength \herschel bands which are crucial to determine the temperature. \cite{magnelli2014} does exclude the bands for which they predict the flux is doubled by the effect of clustering, however this threshold is extreme and will leave substantial clustering signal in their photometry.

To further check for systematic issues in our stacked fluxes, we applied our model to the stacked SEDs of \cite{bethermin2015-a}, which were obtained for a different sample, in a single mass bin, and include longer wavelength data from LABOCA $870\,\um$ and AzTEC $1.1\,\mm$ at all redshifts. The correction for clustering is also performed with another method. Despite these differences, the evolution of $\tdust$ in these data is in excellent agreement with the above relation, suggesting that our stacked fluxes are robustly measured. We also compare the $\mean{U}$ value as reported by \cite{bethermin2015-a} for reference; the \citetalias{draine2007} model used in that work assumes a different functional form for the $U$ distribution, so the comparison is limited in scope, but we nevertheless recover a similar slope for the redshift dependence of the temperature.

We therefore confirm that the dust temperature increases continuously from about $25\,\kelvin$ at $z=0$ up to more than $40\,\kelvin$ at $z=4$, with little to no dependence on stellar mass (hence $\lir$) at fixed redshift. This is important to take into account when only limited FIR photometry is available and $\tdust$ needs to be assumed to extrapolate the total IR luminosity, particularly for sub-millimeter samples which do not probe the emission blueward of the peak of the dust emission (see \rsec{SEC:irsed_mono}).

\subsubsection{Evolution of the IR8}

\cite{elbaz2011} proposed that a unique value of $\ireight=4.9$ holds for all main sequence galaxies, however it could be seen already from their stacked data (see their Fig.~7) that the average $\ireight$ is closer to $8$ at $z=2$ (see also \citealt{reddy2012}).

After applying the correction described in \rsec{SEC:stack_correct}, we found $\ireight \sim 7$ at $z=2$, mildly but significantly larger than the value first reported in \cite{elbaz2011} at $z=1$. On the other hand, we found the $z=0$ galaxies from the HRS have a marginally smaller $\ireight$ of about $3.5$, very similar to our $z=1$ value of $4$. This implies that the average value has evolved continuously between $z=2$ and $z=1$ only, and our stacks at $z=3$ suggest that this evolution stops at higher redshifts. We thus fit the evolution of the average $\ireight$ as a broken linear relation and obtained
\begin{equation}
\ireight^{\rm MS} = 4.08\pm0.29 + (3.29\pm0.24) \times \left\{\begin{array}{lcl}
0 & \text{if} & z < 1\,, \\
(z-1) & \text{if} & 1 \leq z \leq 2\,, \\
1 & \text{if} & z > 2\,.  \label{EQ:ir8_ms}
\end{array}\right.
\end{equation}
This relation is displayed in \rfig{FIG:tdust_fpah_stack} (bottom right) and compared to the values obtained by \citet[Fig.~7]{elbaz2011} and \cite{magdis2012}, which are in rough agreement. The $\ireight$ value of \cite{magdis2012} at $z=0$ is significantly higher than ours, probably because their local sample was flux-limited \citep{dacunha2010-a}, hence is biased toward starbursts which have a systematically higher $\ireight$ (see \rsec{SEC:selection_effects}).

Interestingly, we found that low mass galaxies ($\mstar < 10^{10}\,\msun$) have systematically higher $\ireight$ values, implying that their PAH emission is reduced. \cite{shivaei2017} observed a similar trend in $z\sim2$ galaxies with spectroscopic redshifts. Shivaei et al.~also observe that this trend of reduced PAH emission can be linked to decreasing metallicity, as observed in the local Universe \citep[e.g.,][]{galliano2003,ciesla2014}. The physical origin of this trend is still debated. One plausible explanation is that a metal-poor interstellar medium blocks less efficiently the UV radiation of young stars, and makes it harder for PAH molecules to survive \citep[e.g.,][]{galliano2003}. Other scenarios have been put forward, suggesting either that low metallicity objects are simply too young to host enough carbon grains to form PAH complexes \citep{galliano2008}, or that this is instead caused by a different filling factor of molecular clouds in metal poor environments \citep{sandstrom2012}. Metallicity, in turn, is positively correlated with the stellar mass through the mass-metallicity relation \citep{lequeux1979,tremonti2004}, and this relation has been found to evolve with time, so that galaxies were more metal-poor in the past \citep[e.g.,][]{erb2006}. If metallicity was the main driver of PAH emission, one would expect to find the strongest PAH features (and the lowest $\ireight$) within massive low-redshift galaxies, which is indeed what we found.

\begin{figure}
    \centering
    \includegraphics[width=0.48\textwidth]{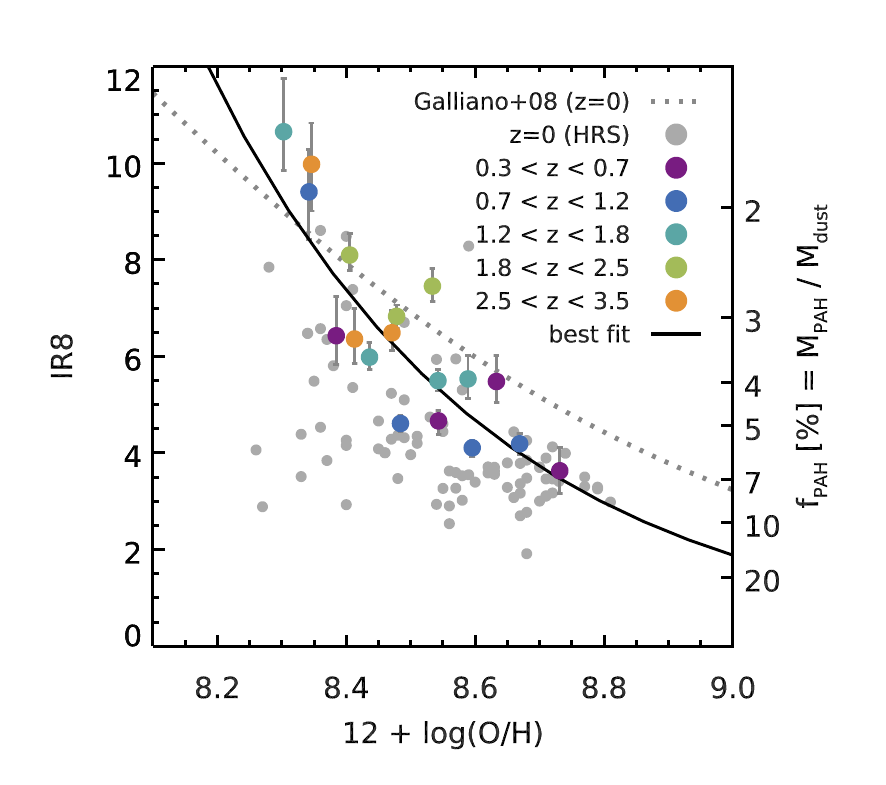}
    \caption{Relation between the $\ireight$ observed in stacked \spitzer and \herschel photometry, and the gas-phase metallicity. The metallicity, expressed in terms of oxygen abundance $12+\log_{10}(O/H)$ (the solar value is $8.73$, \citealt{asplund2009}), was estimated using the Fundamental Metallicity Relation \citep[FMR,][]{mannucci2010}. Data points are colored with redshift as indicated in the legend. We also show the galaxies from the HRS as gray circles, using their measured metallicities. The black line shows the best fitting power law to the stacked data. The $z=0$ relation obtained by \cite{galliano2008} is shown for reference with a dotted gray line, converting their measured PAH mass fractions to $\ireight$ using \req{EQ:fpah_ir8}.}
    \label{FIG:fpah_metal}
\end{figure}

To test this hypothesis quantitatively, we used the Fundamental Metallicity Relation (FMR; \citealt{mannucci2010}) to estimate the average metallicity of our stacked galaxies from their stellar masses and SFRs. The resulting relation between metallicity and $\ireight$ is shown in \rfig{FIG:fpah_metal}. We found a clear anti-correlation between the two quantities, quantitatively matching the trend observed in the local Universe by \cite{galliano2008} and confirming the results of \cite{shivaei2017}. The best fitting power law is
\begin{equation}
\ireight = (3.5 \pm 0.3)\times(Z/Z_\sun)^{-0.99\pm0.15}\,,
\end{equation}
where $Z$ is the metallicity (which can be substituted for the oxygen abundance $O/H$ under the assumption that $O/H$ scales linearly with $Z$). The galaxies of the HRS, for which we have individual metallicity measurements \citep{boselli2010}, also follow a similar trend.

Unfortunately, the power of the FMR in predicting metallicities is limited \citep[particularly at high redshifts, e.g.,][]{bethermin2015-a}, and measuring metallicities from emission lines for individual galaxies is both expensive and prone to systematics \citep{kewley2008}. Excluding the masses below $10^{10}\,\msun$, the model of \req{EQ:ir8_ms} with a simple dependence on redshift provides a fit of equal quality to the data. Since \req{EQ:ir8_ms} is more easily applicable to large samples, we chose to adopt it as the fiducial model and caution that the relations derived below for $\ireight$ only apply to galaxies more massive than $10^{10}\,\msun$.

\subsection{Values for individual galaxies \label{SEC:irsed_indiv}}

In this section we describe the scatter on both $\tdust$ and $\ireight$ about the average ``main sequence'' values we obtained in \rsec{SEC:irsed_stack}. We also describe how these quantities are modified for starburst galaxies, that is, those galaxies that have an excess $\sfr$ at a given stellar mass compared to the main sequence. To quantify this latter excess, we used the ``starburstiness'' \citep{elbaz2011} which is defined as $\rsb \equiv \sfr/\sfrms$; galaxies with $\rsb=1$ are on the main sequence, and those with $\rsb>1$ are located above the sequence.

\subsubsection{Fitting individual galaxies}

\begin{figure*}
    \centering
    \includegraphics[width=0.9\textwidth]{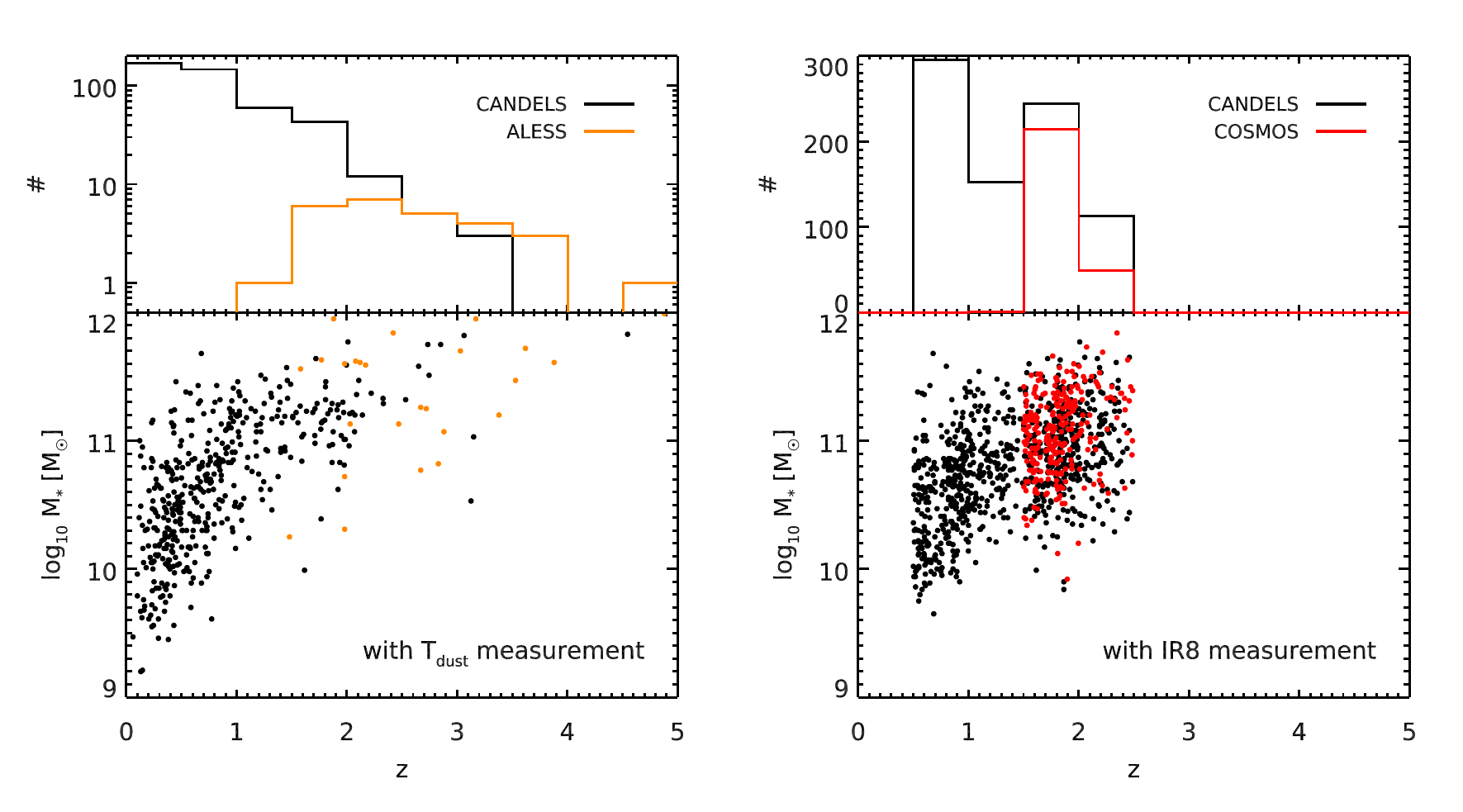}
    \caption{Distribution of galaxies with a $\tdust$ (left) or $\ireight$ (right) measurement in the $\mstar$-redshift plane. The galaxies from CANDELS are shown with black circles, that from the larger COSMOS field in red, and the smaller sample of ALESS galaxies is shown in orange. At the top we display the redshift distribution of each sample.}
    \label{FIG:indiv_sample}
\end{figure*}

We used our library to fit the FIR-detected galaxies in the CANDELS fields and COSMOS $2\,\sq\degr$. To ensure reliable fits, we selected galaxies with a good enough wavelength coverage and robust photometry. In particular we only used the \herschel photometry for clean sources \citep{elbaz2011} to avoid biases toward low $\tdust$ because of blending, and excluded all sources for which the matching of the $24\,\um$ emission to a $H$ or \Ks-band counterpart was ambiguous \citep[as defined in][]{schreiber2016}. We also excluded fits of poor quality by rejecting galaxies with $\chi^2$ larger than $10$ (less than $10\%$ of our sample), indicative of uncaught issues in the photometry and counterpart identification.

For the measurement of $\tdust$, following \cite{hwang2010-a} we required at least one photometric measurement with significance greater than $5\sigma$ on either side of the peak of the dust continuum. This produced a sample of $438$ galaxies, which are shown on the left panel of \rfig{FIG:indiv_sample}.

For the measurement of $\ireight$, we selected only the galaxies at $0.5<z<1.5$ with at least a $3\sigma$ detection in \spitzer IRS $16\,\um$ and the galaxies at $1.5<z<2.5$ with at least a $3\sigma$ detection in \spitzer MIPS $24\,\um$ to ensure the rest-frame $8\,\um$ emission is well constrained. Of these, we only kept the galaxies for which $\lir$ could be independently measured using longer wavelength photometry at better than $5\sigma$. This produced a sample of $1068$ galaxies, shown on the right panel of \rfig{FIG:indiv_sample}, and $264$ of these are in COSMOS $2\,\sq\degr$.

\subsubsection{Completeness and selection biases \label{SEC:selection_effects}}

To quantify the selection biases introduced by the numerous criteria listed above, we created a new mock catalog with EGG (see \rsec{SEC:stack_correct}) over $10\deg^2$ and down to a \spitzer IRAC $3.6\,\um$ magnitude of $25$ (slightly deeper than the typical $5\sigma$ depth in CANDELS). We then perturbed the generated fluxes within the uncertainties typical for CANDELS, and mimicked the \spitzer and \herschel flux extraction procedure described in \cite{elbaz2011}: we only kept the $16$ and $24\,\um$ fluxes for galaxies with a $3.6\,\um$ flux larger than $0.5\,\um$ \citep{magnelli2011}, we only kept the $70$ and $100\,\um$ fluxes for galaxies with a $24\,\um$ flux larger than $21\,\uJy$ ($3\sigma)$, we only kept the $160$ to $500\,\um$ fluxes for galaxies with a $24\,\um$ flux larger than $35\,\uJy$ ($5\sigma$), and finally we only kept the $350$ and $500\,\um$ fluxes for galaxies with a $250\,\um$ flux larger than $3.8\,\mJy$ ($2\sigma$). Since most of our sample has no ALMA data, we did not include ALMA photometry in the mock catalog.

We then ran the same fitting procedure as for the real galaxies, and applied the same selection criteria to produce the two samples with robust $\tdust$ and $\ireight$. Normalizing the number of objects in the mock catalog over an area equal to that of CANDELS (and accounting for the loss of effective area caused by the rejection of the non-clean \herschel sources), the mock catalog predicts $396$ galaxies with a robust $\tdust$ measurement, and $1101$ galaxies with a robust $\ireight$. These numbers are in excellent agreement with the observed ones, and confirm the validity of the mock catalog.

\begin{figure*}
    \centering
    \includegraphics[width=\textwidth]{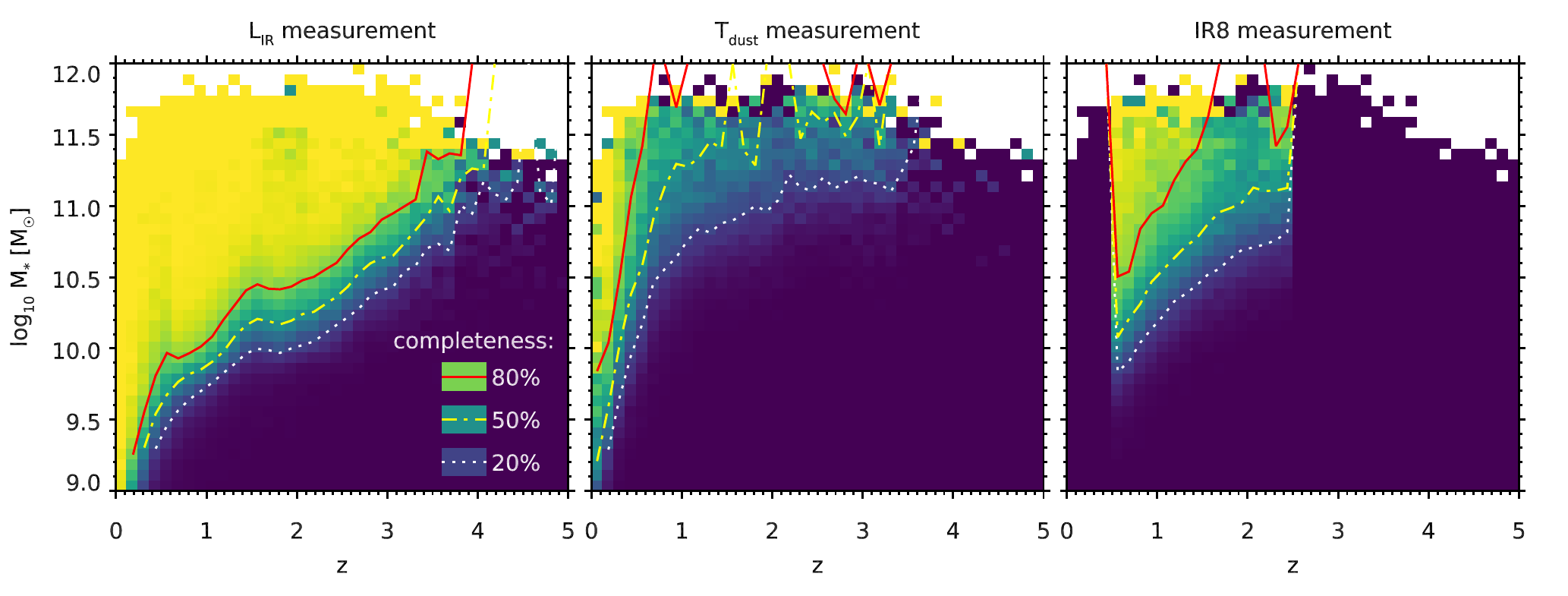}
    \caption{{\bf Left:} Completeness of a CANDELS-like sample computed from the mock catalog, limited to galaxies with a measurement of $\lir$ with $S/N > 3$, from either \spitzer MIPS or \herschel. The completeness is shown as a function of redshift and stellar mass: each cell of the plot displays the completeness is the corresponding bin of redshift and mass. The $80$, $50$ and $20\%$ completeness levels are indicated with red, yellow and white lines, respectively. {\bf Center:} Same as left, but limited to the sample with a robust $\tdust$ measurement. {\bf Right:} Same as left, but limited to the sample with a robust $\ireight$ measurement.}
    \label{FIG:mz_comp}
\end{figure*}

We display in \rfig{FIG:mz_comp} the completeness of various samples: selecting galaxies with an $\lir$ measured at better than $3\sigma$, selecting galaxies with a robust $\tdust$, and selecting galaxies with a robust $\ireight$. It is clear from this plot that one can only obtain robust measurements of $\tdust$ or $\ireight$ for a fraction of the IR-detected galaxies. For the dust temperature, our sample reaches $80\%$ completeness only at $z<0.6$, however it can still probe galaxies with $\mstar > 10^{11}\,\msun$ at much higher redshifts (up to $z\sim$ 3) albeit with a lower completeness of $20\%$; this implies that selection effects are important and have to be studied carefully. The situation for the $\ireight$ is not as dramatic, as the sample reaches $80\%$ completeness at $z=1$ above $\mstar > 8\times10^{10}\,\msun$, and $50\%$ completeness at $z=2$ above the same mass.

\begin{figure*}
    \centering
    \includegraphics[width=\textwidth]{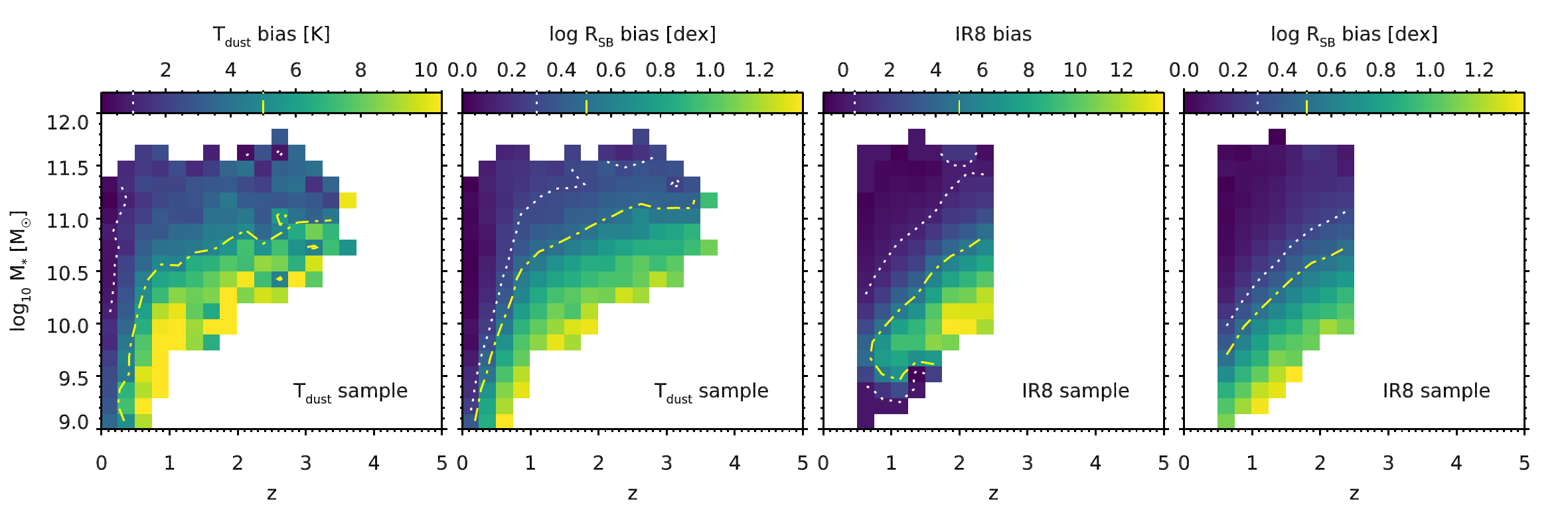}
    \caption{{\bf Left:} Bias in the average $\tdust$ as a function of redshift and mass, computed from the mock catalog and for the ``robust $\tdust$'' sample. Each cell of the plot shows the difference between the observed average $\tdust$ in the mock and the true average of the mock in the corresponding bin of redshift and mass. Biases of $1$ and $5\kelvin$ are shown with white and yellow lines, respectively. {\bf Center-left:} Same plot, but showing instead the bias in ``starburstiness'' ($\rsb$), or offset from the main sequence. Biases of a factor two and three are shown with white and yellow lines, respectively. {\bf Center-right:} Same as left, but showing the bias in $\ireight$ for the ``robust $\ireight$'' sample. Biases of $0.5$ and $5$ are shown with white and yellow lines, respectively. {\bf Right:} Same as center-left, but for the ``robust $\ireight$'' sample.}
    \label{FIG:mz_bias}
\end{figure*}

We display in \rfig{FIG:mz_bias} the resulting selection biases on $\tdust$, $\ireight$ and $\rsb$. We found that our selection criteria bias our samples toward galaxies with higher $\sfr$ at fixed mass, hence to galaxies offset from the main sequence. For example, our sample with $\tdust$ measurement will only probe galaxies a factor three above the main sequence for masses less than $10^{11}\,\msun$ at $1<z<3$. Because $\tdust$ and $\ireight$ both correlate with $\rsb$ \citep{elbaz2011,magnelli2014}, this translates into a positive bias on both quantities: at a given mass and redshift, the observed average $\tdust$ and $\ireight$ of our sample are higher than their true average. This is particularly the case for $\tdust$, and is apparent also on the real data set (\rfig{FIG:indiv}). Interestingly, although our $\tdust$ measurements require a detection in the SPIRE bands, which could bias our sample toward colder temperatures, we find that this is a negligible effect compared to the bias toward starbursts: on no occasion is the average measured $\tdust$ lower than the true average.

This analysis implies that the bias toward higher $\rsb$ is the dominant source of bias on the measured dust properties. In the following, we focus on the trend of both $\tdust$ and $\ireight$ with $\rsb$, and therefore we will not be affected by this bias. Yet it is important to keep in mind that our results are based almost entirely on galaxies with masses larger than $3\times10^{10}\,\msun$; studying galaxies at lower masses will require next generation instruments such as ALMA and \jwst.

\subsubsection{Dust temperatures \label{SEC:indiv_tdust_result}}

\begin{figure*}
    \centering
    \includegraphics[width=\textwidth]{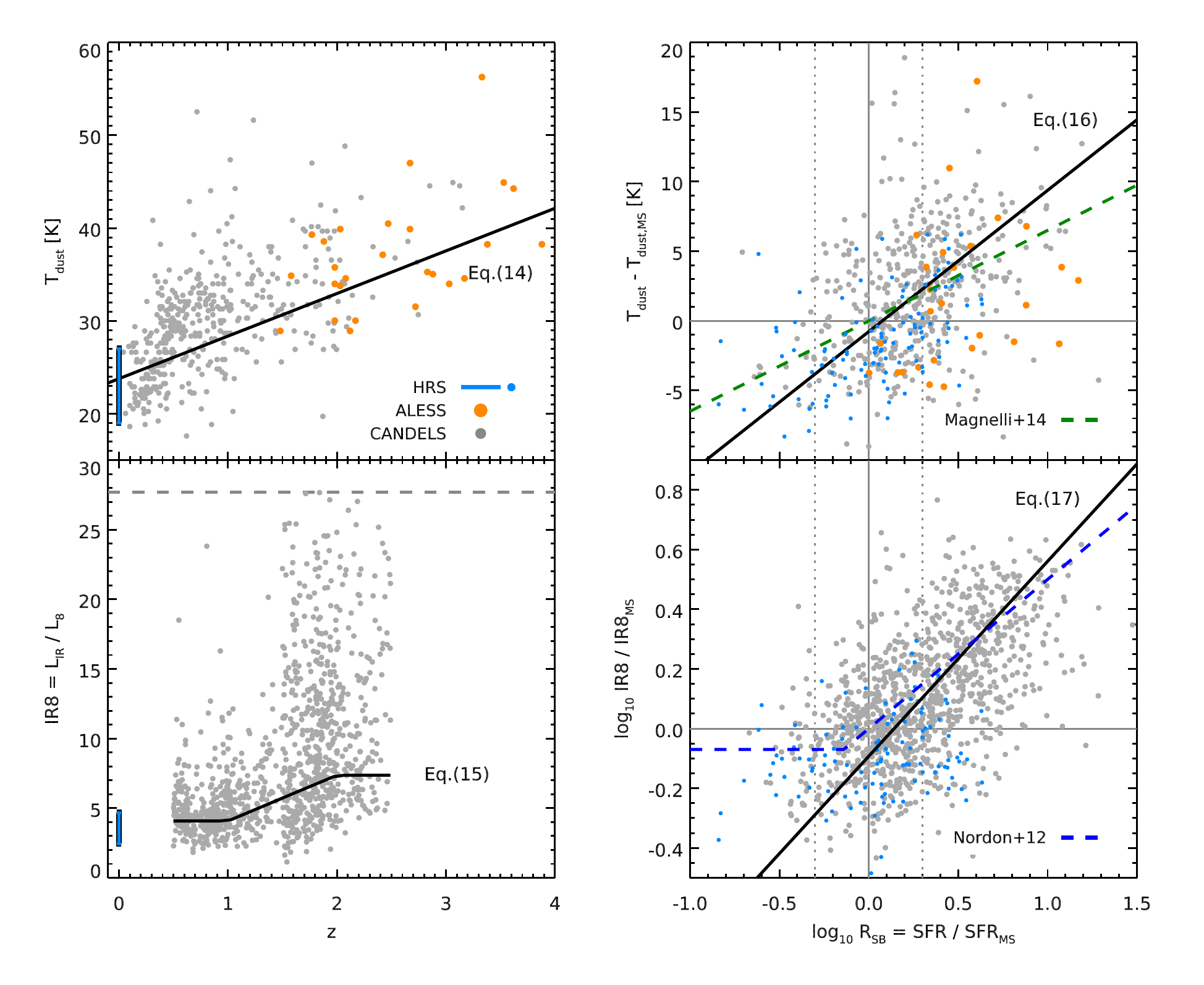}
    \caption{{\bf Left:} Evolution of the dust temperature ($\tdust$, top) and $\ireight\equiv\lir/\leight$ (bottom) of galaxies individually detected with \herschel in the CANDELS fields (gray dots), from ALESS (orange circles) and the HRS (blue circles or range). We overplot the trends found in stacking (\rsec{SEC:irsed_stack}) with solid black lines. The dashed horizontal line indicates the maximum $\ireight$ value that our library can reach. {\bf Right:} Evolution of $\tdust$ (top) and $\ireight$ (bottom) with the starburstiness ($\rsb$, see text). The legend is the same as for the plots on the left, except that here the black solid line shows our best-fit relation to the data. For $\tdust$ we show the relation previously found by \cite{magnelli2014} with a dashed green line, and for $\ireight$ we show the relation of \cite{nordon2012} with a dashed blue line.}
    \label{FIG:indiv}
\end{figure*}

In \rfig{FIG:indiv} (top left) we show the measured dust temperatures for individual \herschel detections, ALESS galaxies, and galaxies from the HRS. These temperatures match broadly the trend observed in the stacked SEDs, but with a tendency to show systematically larger values. As discussed in the previous section, this is to be expected: the requirement of a well-measured IR SED biases this sample towards strongly star-forming starburst galaxies (particularly at high redshifts, $z>0.5$), which have higher temperatures. Subtracting our redshift-dependent average from the measured $\tdust$ values (\req{EQ:tdust_ms}), we then observed how the residuals correlate with the offset from the main sequence. The result is shown in \rfig{FIG:indiv} (top right). We found a positive correlation and parametrized it with a linear relation (obtained as the bisector of the data):
\begin{equation}
\tdust [\kelvin] = \tdust^{MS} - (0.77 \pm 0.04) + (10.1\pm0.6) \times \log_{10}(\rsb) \,. \label{EQ:tdust_sb}
\end{equation}
where $\tdust^{MS}$ is defined in \req{EQ:tdust_ms}. The error bars were determined by bootstrapping. Such a trend was first observed in \cite{elbaz2011}, and later quantified by \cite{magnelli2014} who stacked galaxies at various locations on the $\sfr$--$\mstar$ plane (see also \citealt{matsuki2017}). They reported a linear relation between $\tdust$ and $\log_{10}(\rsb)$ --- which they call $\Delta\log(\ssfr)$ --- with a slope of $6.5\,\kelvin$, which is shallower than the one we measure here but still roughly fits our data.

The residual intrinsic scatter of the temperatures is presented in \rfig{FIG:tdust_scatter} as a function of redshift, after subtracting the measurement and redshift uncertainties (assuming $\Delta z/(1+z) = 3\%$; \citealt{pannella2015}). The absolute scatter (in Kelvins) increases mildly with redshift, and this evolution is consistent with a constant relative scatter of $12\%$, relative to $\tdust^{\rm MS}$ as given in \req{EQ:tdust_ms}. The evolution of this scatter beyond $z=2$ is poorly constrained, and larger samples with $870$ or $1200\,\um$ coverage would be required to determine whether it keeps increasing at higher redshifts. Over the whole sample, the scatter in temperature was initially $17\%$ (relative to the median). After removing the evolution of the main sequence temperature with redshift, this scatter dropped to $15\%$ (relative to $\tdust^{\rm MS}$), and removing the starburstiness trend further reduced the scatter to the final value of $12\%$.

\subsubsection{Temperature -- luminosity relation \label{SEC:lir_tdust}}

\begin{figure*}
    \centering
    \includegraphics[width=0.9\textwidth]{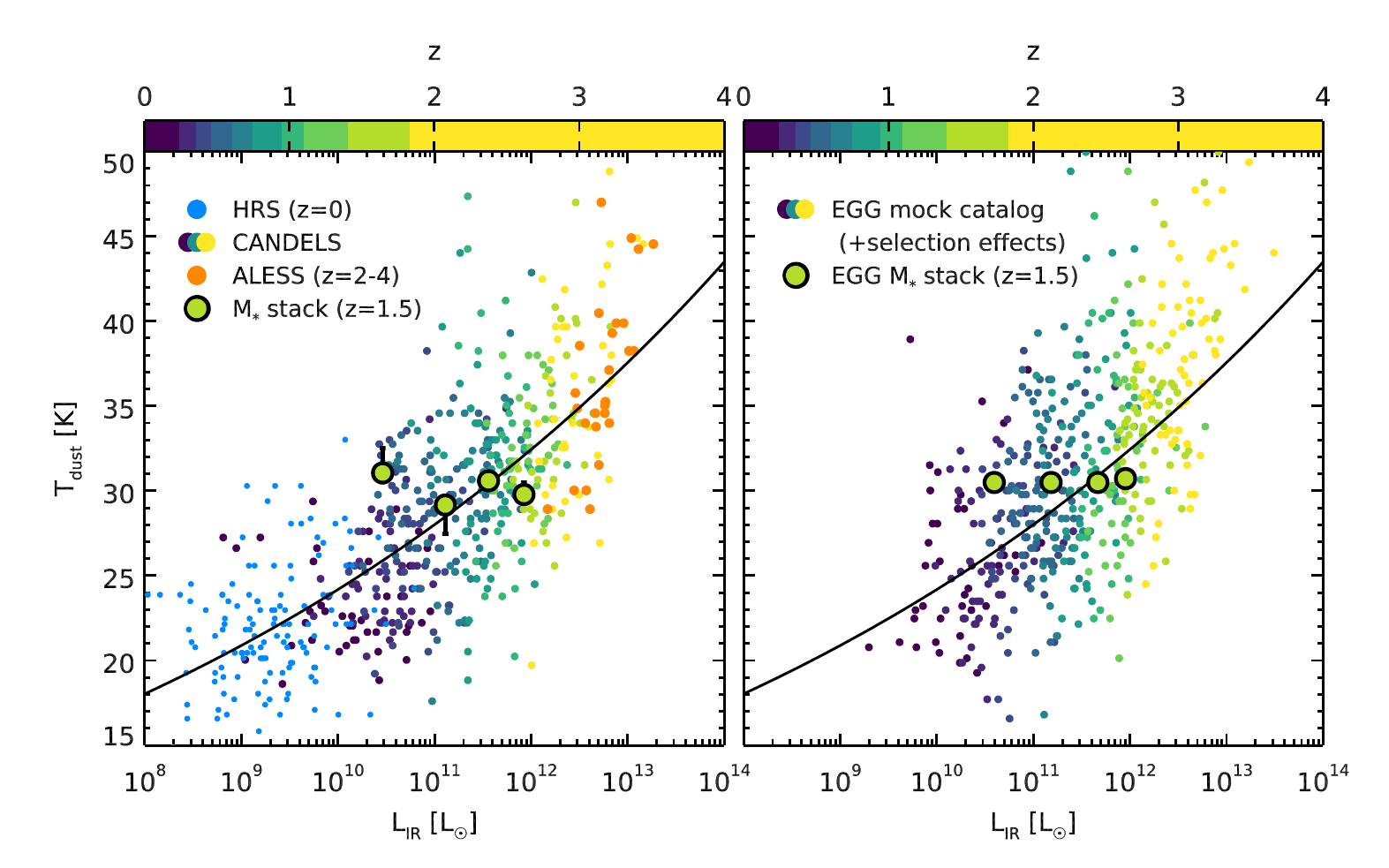}
    \caption{{\bf Left:}Relation between the dust temperature ($\tdust$) and the total infrared luminosity ($\lir$) for galaxies individually detected with \herschel in the CANDELS fields (colored circles, dark purple to yellow from low to high redshift), from ALESS (orange circles) and the HRS (light blue circles). We overplot the best-fit power law (see text) with a black solid line, as well as the values obtained at $z=1.5$ by stacking galaxies in different bins of mass (see \rsec{SEC:irsed_stack}). {\bf Right:} same as left, but for the mock CANDELS catalog produced with EGG (defined in \rsec{SEC:selection_effects}).}
    \label{FIG:lir_tdust}
\end{figure*}

Historically, the first studied correlation was that between the dust temperature and the infrared luminosity (see references in \rsec{SEC:introduction}, \S7). Whether this correlation or that involving the redshift and the starburstiness provides the best description of the observations has been repeatedly debated in the literature \citep[see, e.g.,][]{casey2012-a,symeonidis2013,magnelli2014}. Unfortunately, owing to the strong selection effects of flux-limited FIR samples, it is generally difficult to de-correlate the effect of redshift and luminosity. As shown in \rfig{FIG:lir_tdust} (left), our sample is affected by this issue: while our galaxies do show a clear correlation between $\tdust$ and $\lir$, with $\tdust = 5.57\times\lir^{0.0638}$, $\lir$ also strongly correlates with the redshift. The residual scatter around the $\lir$--$\tdust$ relation is $13\%$, which is comparable to that found in the previous section; individual galaxies do not favor one description over the other.

At present, only stacking has enough discriminative power to address this question. For example, \cite{magnelli2014} found, at fixed redshift, a tighter correlation between $\tdust$ and $\rsb$ than with $\lir$ in their stacks. In our stacks in bins of stellar mass and redshift, we found no evidence for a correlation between $\tdust$ and $\lir$ at fixed redshift; for example at $z=1.5$, $\tdust$ remains mostly constant while $\lir$ spans almost two orders of magnitude, as shown in \rfig{FIG:lir_tdust} (left). This suggests that there exists no fundamental correlation between $\lir$ and $\tdust$ (at a given redshift) beyond that induced by the starburstiness, which gets averaged-out in our stacks, and selection effects.

To further demonstrate this, we show in \rfig{FIG:lir_tdust} (right) the $\lir$--$\tdust$ relation arising in the mock catalog produced with EGG (see \rsec{SEC:selection_effects}). This relation is very similar to that observed in the real catalog, albeit with a slightly higher scatter (15\%). This $\lir$--$\tdust$ relation was not imposed when generating the mock, instead it emerges naturally as a combination of selection effects and the relations discussed in the previous section.

\subsubsection{IR8 \label{SEC:indiv_ir8_result}}

We applied the same procedure to $\ireight$, and the results are shown in \rfig{FIG:indiv} (bottom left and right). Consistently with the results of \cite{elbaz2011} and \cite{nordon2012}, we found a correlation between $\ireight$ and $\log_{10}(\rsb)$, meaning that starburst galaxies have depressed PAH emission, which Elbaz et al.~interpreted as a sign of increased compactness of the star-forming regions. We modeled this dependence with a linear relation (obtained as the bisector of the data):
\begin{equation}
\ireight = \ireight^{\rm MS} \times (0.81 \pm 0.02) \times \rsb^{0.66 \pm 0.05}\,.\label{EQ:ir8_sb}
\end{equation}
This relation is consistent with that derived by \cite{nordon2012}, although we do not find the need for a separate regime for galaxies below the main sequence. Over the whole sample, the scatter in $\log_{10}(\ireight)$ is $0.28\,\dex$. This is reduced to $0.22$ and $0.18\,\dex$ after subtracting the redshift and starburstiness dependences, respectively.

\begin{figure}
    \centering
    \includegraphics[width=0.48\textwidth]{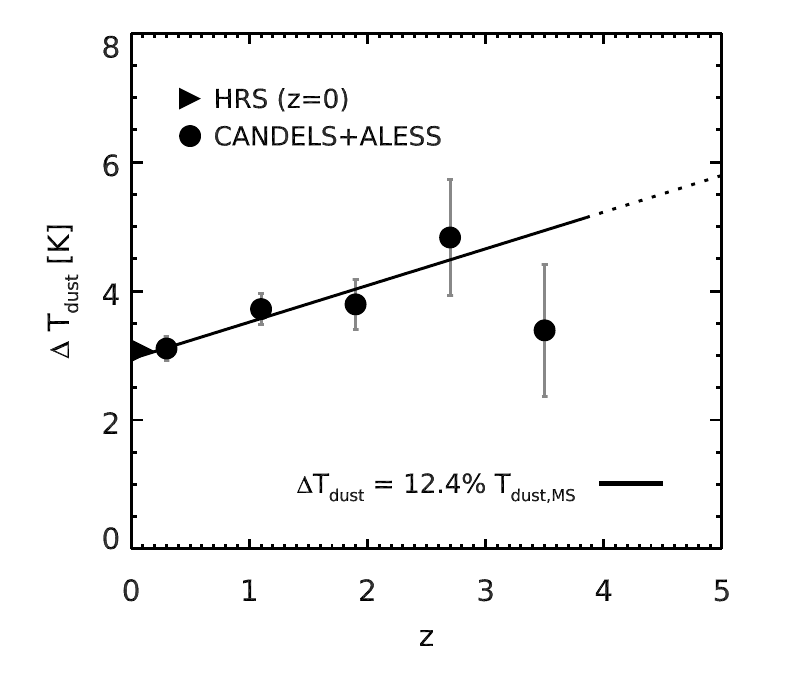}
    \caption{Residual intrinsic scatter of $\tdust$, obtained after removing the redshift evolution (\req{EQ:tdust_ms}) and starburstiness dependence (\req{EQ:tdust_sb}), and subtracting statistically the uncertainty on the $\tdust$ measurements and on the redshift. Measurements of the scatter in CANDELS and ALESS are shown with solid circles, and the HRS is shown with a solid triangle. The trend with redshift is modeled as a constant relative uncertainty (black line).}
    \label{FIG:tdust_scatter}
\end{figure}

\section{Optimal $\lir$ and $\mdust$ measurements \label{SEC:irsed_recipe}}

Contrary to the standard FIR libraries from the literature \citepalias[e.g.,][]{chary2001}, ours has three degrees of freedom: the normalization (either $\lir$ or $\mdust$), the dust temperature, and the $\ireight$ (or $\fpah$). This requires particular care when the observed SEDs are poor and some of these parameters are unconstrained. To obtain a robust determination of $\lir$ for galaxies with variable wavelength coverage, the procedure we recommend is described in the following subsections. In the next section we quantify the accuracy of monochromatic $\lir$ or $\mdust$ measurements (i.e, when only a single broadband flux is available for a galaxy) when this procedure is applied, and make predictions for \jwst and ALMA.

\subsection{Selection of the free parameters}

The dust temperature must be fixed if the peak of the FIR emission is not constrained. In practice, this requires at least one measurement at more than $3\sigma$ on either side of the peak \citep{hwang2010-a}, and within the rest-frame $15\,\um$ to $3\,{\rm mm}$ to avoid contribution from PAH or free-free emission. Here we defined the ``peak'' by first fitting the galaxy with a free $\tdust$, and measuring $\lambda_{\rm max}$ from the resulting best-fit template. To fix the temperature, one will use \req{EQ:tdust_ms} evaluated at the redshift of the galaxy. The library is then reduced to a single dust continuum and a single PAH template.

Likewise, the $\ireight$ must be fixed if no measurement probes the rest-frame $5$ to $15\,\um$ (to constrain $\leight$), or if no observation is available for wavelengths greater than $15\,\um$ (to constrain $\lir$). In this case the value of $\ireight$ should be taken from \req{EQ:ir8_ms} and evaluated at the redshift of the galaxy. Using \req{EQ:fpah_ir8_th}, $\ireight$ can then be translated to $\fpah$ and fix the relative amplitude of the PAH and continuum templates.

\subsection{Fitting the observed photometry}

For each value of $\tdust$ allowed in the fit, the templates (provided in rest-frame quantities) must be translated to the observer frame and integrated under the filter response curves of the available photometry. Using a linear solver, one can then fit the convolved templates to the observed photometry as a linear combination of the dust continuum and PAH emission:
\begin{equation}
S_{\nu}^{\rm model} = \mdust^{\rm cont}\,S_{\nu}^{\rm cont} + \mdust^{\rm PAH}\,S_{\nu}^{\rm PAH}\,, \label{EQ:fit_model_full}
\end{equation}
where $\mdust^{\rm cont}$ and $\mdust^{\rm PAH}$ are free parameters, and compute the $\chi^2$. Among all the templates in the library, one can then pick as the best-fit solution the $\tdust$ value which produced the smallest $\chi^2$ (or the only available $\tdust$ value if it is kept fixed). If the fit is performed with a standard linear solver, the amplitude of either component can become negative. This typically happens when the observed photometry requires an $\ireight$ value larger than what the library has to offer (which is rare by construction), and the fit uses a PAH component with a negative amplitude to reduce the mid-IR emission. In such cases, one could fix $\fpah=0$ (i.e., no PAH emission) or exclude the rest-frame $8\,\um$ photometry altogether. Such situations can hint at the presence of obscured AGNs \citep{donley2012}, or galaxies with strong silicate absorption \citep{magdis2011-a}.

If $\ireight$ is fixed, the two components of the library are merged into one single template for each value of $\tdust$. \req{EQ:fit_model_full} becomes
\begin{equation}
S_{\nu}^{\rm model} = \mdust\,\Big[(1-\fpah) \times S_{\nu}^{\rm cont} + \fpah \times S_{\nu}^{\rm PAH}\Big]\,,
\end{equation}
where $\fpah$ is computed from $\ireight$ using \req{EQ:fpah_ir8_th}, and the only free parameter is $\mdust$. Similarly to the procedure above, one can then vary $\tdust$ and choose as the best-fit solution the value of $\tdust$ that produced the smallest $\chi^2$.

The dust mass is computed as $\mdust = \mdust^{\rm cont} + \mdust^{\rm PAH}$, while the other parameters ($\lir$ and $\leight$) can be obtained from the tabulated values corresponding to the best-fit $\tdust$.

\subsection{Computing uncertainties on best-fit parameters}

The most straightforward and secure way to determine uncertainties is to perform a Monte Carlo simulation for each galaxy, where the observed fluxes are randomly perturbed with a Gaussian scatter of amplitude set by the flux uncertainties. This needs to be repeated at least $100$ times for reliable $1\sigma$ error bars. For each parameter, the probability distribution function can be determined from the distribution of best fit values among all random realizations.

\section{Accuracy of monochromatic measurements \label{SEC:irsed_mono}}

\begin{figure*}
    \centering
    \includegraphics[width=\textwidth]{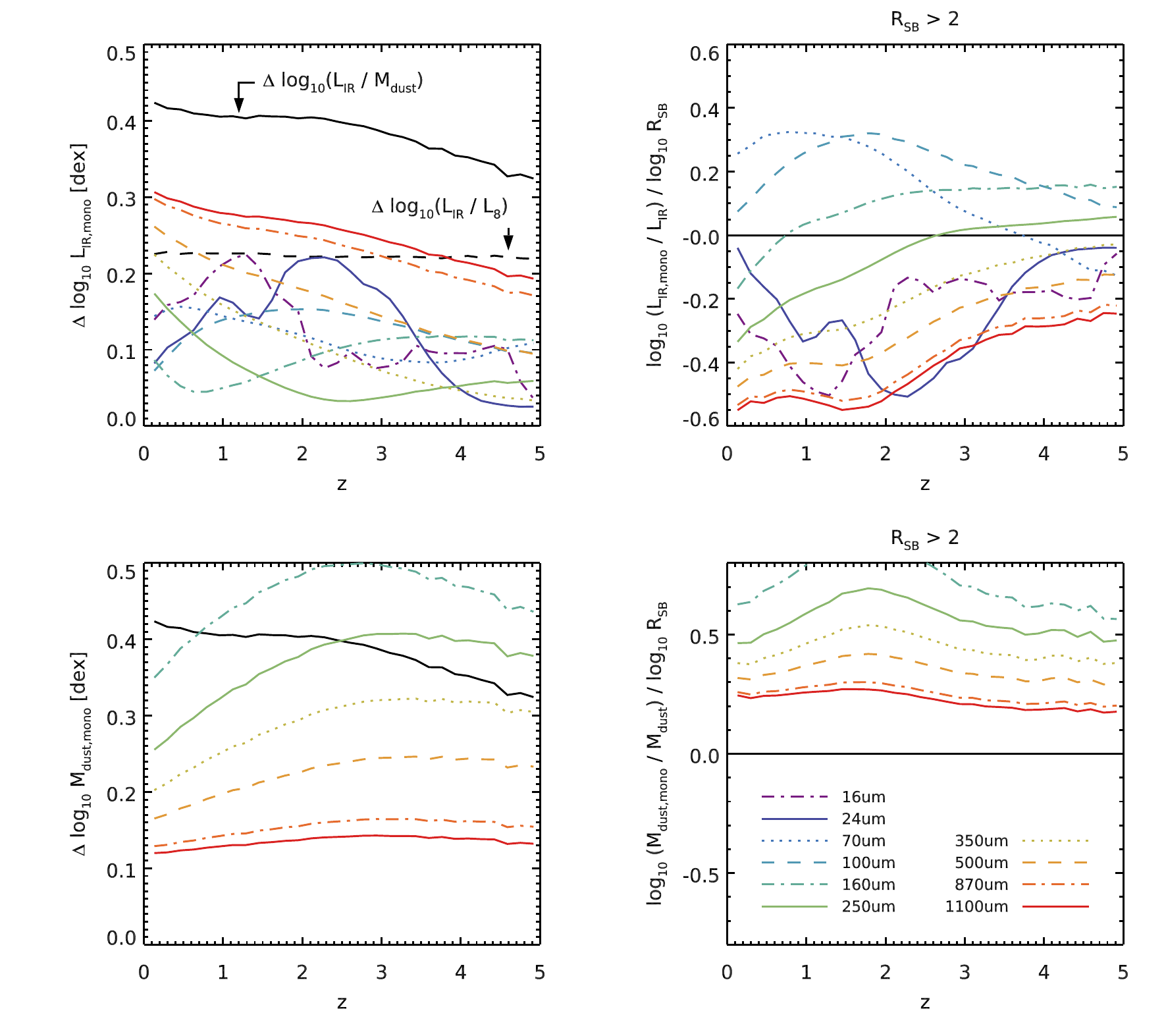}
    \caption{{\bf Top left:} Predicted evolution of the uncertainty in $\lir^{\rm mono}$, that is, the $\lir$ inferred from a single broadband photometric measurement, each band corresponding to a different color and line style (see legend). This uncertainty is derived by measuring the standard deviation of the difference between the true $\lir$ that was put in the simulated catalog and the observed monochromatic rest-frame luminosity. This is an optimal uncertainty, assuming 1) no error on the measured flux, 2) knowledge of the best average $\lir/L_{\nu}$ conversion factor (i.e., the best average SED), 3) perfect subtraction of the stellar component (which only matters for $16$ and $24\,\um$ at high-redshift), and 4) no contamination from AGNs. For comparison, we show with a black solid and dashed lines the logarithmic scatter in $\lir/\mdust$ and $\lir/\leight$ in the sample at each redshift, which are the main drivers of SED variations at $\lambda > 30$ and $\lambda < 30\,\um$ in our model, respectively. {\bf Top right:} Predicted systematic error on the $\lir$ of starburst galaxies (selected here with $\rsb > 2$) normalized to the galaxies' offset from the main sequence. In other words, a value of $x$ on this plot means that the $\lir$ will be wrong on average by a factor $\rsb^{x}$. {\bf Bottom left \& right:} Same as top, but for $\mdust^{\rm mono}$.}
    \label{FIG:pred_disp}
\end{figure*}

While our library provides three degrees of freedom, in most cases the observed SED of a galaxy will consist of only a couple of data points. The procedure we described in the previous section provides a simplified fit procedure where the number of degrees of freedom is progressively reduced until only one is left: the normalization of the template SED. At fixed $\tdust$, this normalization can be translated either into $\lir$ or $\mdust$, which are usually the quantities observers are after. Often, only one observed data point is available --- typically either from \spitzer MIPS $24\,\um$, ALMA $870\,\um$, or $1.1\,{\rm mm}$, but also in the future from \textit{James Webb}. In this case we dub the measurement of $\lir$ or $\mdust$ as ``monochromatic''.

In the case of such monochromatic measurements, since the shape of the SED depends on $\tdust$ in a strongly non-linear way, the uncertainty on $\tdust$ will propagate into different uncertainties on $\lir$ and $\mdust$ depending on which band is used to determine the normalization of the template (the case of the many ALMA bands in the millimeter regime is discussed in \rsec{SEC:alma_mdust_lir}). We have shown in \rsec{SEC:indiv_tdust_result} that fixing $\tdust$ to the average value for galaxies at a given redshift (\req{EQ:tdust_ms}) provides the correct temperature within $15\%$. In this section, we discuss the uncertainties on $\lir$ and $\mdust$ resulting from this uncertainty on $\tdust$, and show which bands are best suited for monochromatic measurements or either quantities. We also discuss how well the rest-frame $8\,\um$ luminosity can be used to measure $\lir$, which is a situation arising for $z\sim1$ to $2$ galaxies too faint to be detected with \herschel but seen by \spitzer MIPS at $24\,\um$ (or in the future with \jwst, which is discussed more thoroughly in \rsecs{SEC:conv} and \ref{SEC:delta_z}).

\subsection{Mock catalog}

Using a mock catalog built with EGG (see \rsec{SEC:stack_correct}), we simulated monochromatic measurements and compared them to the true values of $\lir$ and $\mdust$. We produced a mock catalog spanning $10\,\deg^2$ and selected all the star-forming galaxies more massive than $\mstar = 10^{10}\,\msun$. We then created a flux catalog containing the fluxes of all galaxies in the simulation in all \spitzer and \herschel bands, as well as \jwst MIRI and ALMA. For each galaxy in the mock catalog, we built its ``average'' expected SED from \reqs{EQ:tdust_ms} and \ref{EQ:ir8_ms} and used this SED to convert each flux into a value of $\lir$ and $\mdust$. We note that we did not simulate the stellar continuum for this experiment; we assumed it can be properly constrained and subtracted using the shorter wavelengths. This will be particularly important for the $16$ and $24\,\um$ bands at high redshifts. Likewise, the impact of AGNs was also ignored, and these will increase the uncertainty in these same bands at all redshifts \citep[e.g.,][]{mullaney2011}. Therefore the uncertainties we derived should be considered as lower limits.

The resulting uncertainties are shown in \rfig{FIG:pred_disp} as a function of redshift, for all \spitzer and \herschel bands, as well as two ALMA bands for illustration (band 7 at $8770\um$, and band 6 at $1100\,\um$); the case of \jwst and ALMA are discussed further in \rsec{SEC:conv}. Several interesting features come out of this figure and we describe them in the following sections.

\subsubsection{Infrared luminosity \label{SEC:mono_lir}}

When measuring fluxes on the dust continuum, we found $\lir$ is best measured when the photometric measurement is close to the peak of the FIR SED: the optimal uncertainty is of the order of $0.05\,\dex$, using $100\,\um$ at $z<0.2$, $160\,\um$ at $0.2<z<1.5$, $250\,\um$ at $1.5<z<3.9$, and $350\,\um$ at $3.9<z<5$. These correspond respectively to rest-frame wavelengths of $90$, $86$, $68$ and $64\,\um$, which are precisely the peak wavelengths of the dust SED at each redshift (see also \citealt{schreiber2015}, Fig.~9). Rightward of the peak, the uncertainty rises continuously as the rest wavelength increases, since the emission beyond the rest-frame $250\,\um$ is rather tracing the dust mass (see \rsecs{SEC:mono_dust}, \ref{SEC:alma_mdust_lir}, and \citealt{scoville2014}), and fluctuations in the $\lir/\mdust$ ratio (displayed in \rfig{FIG:pred_disp}, top and bottom left) are driven by the adopted scatter in $\tdust$.

Leftward of the peak, the uncertainties rise significantly up to $0.22\,\dex$ when probing the rest-frame $5$ to $10\,\um$, that is, for $16\,\um$ at $0.5<z<2$ and $24\,\um$ at $1.5<z<3$. This, in turn, is caused by variations of $\ireight$ (displayed in \rfig{FIG:pred_disp}, top left), and shows that fluxes in this wavelength domain should be interpreted carefully. Outside of these ranges dominated by PAH emission, our model suggests that the short wavelengths can be excellent tracers of $\lir$, for example with $24\,\um$ reaching the same accuracy as $350\,\um$ at $z>3.9$. This is linked to our assumption of a constant fraction of very small grains. Little data can back up this assumption at present; in the local Universe \cite{chary2001} reported a scatter of $0.15\,\dex$ between $\lir$ and the $12$ or $16\,\um$ luminosities, which is consistent with the values we obtained with our model and suggest the small grain population does not vary strongly from one galaxy to the next. On the other hand, the hotter $\tdust$ and reduced PAH emission observed in distant galaxies suggest that small grains could get destroyed more efficiently at this epoch. In the near future, \jwst will be able to detect distant galaxies at rest wavelengths less than $12\,\um$ and check this assumption. Another source of uncertainty will be the subtraction of the stellar continuum which can start to dominate below $5\,\um$, although this should be less of a problem at high redshifts where the specific SFRs are higher. Last but not least, we caution that this whole discussion is ignoring AGNs, which seem to be very common at least among massive galaxies at $z>3$ \citep{marsan2017}.

Using a single photometric point, and therefore fixing $\ireight$ and $\tdust$ to their redshift average, one will be systematically biased for those galaxies which have unusual $\ireight$ or $\tdust$ values. As shown in \rsec{SEC:irsed_indiv}, this is the case for starburst galaxies. For this reason, we also display on \rfig{FIG:pred_disp} (top right) the predicted value of this systematic bias for galaxies with $\rsb > 2$ (i.e., at least one sigma away from the main sequence). The trend is for measurements leftward of the peak to barely overestimate the $\lir$ by no more than $\rsb^{0.3}$. In other words, a galaxy which is truly a factor five above the main sequence will be observed instead at a factor eight. On the other hand, measurements rightward of the peak or those dominated by PAH emission can reach systematic underestimations by a factor of $\rsb^{0.5}$, so a galaxy a factor five above the MS will be seen at only a factor $2.2$. These systematic errors will tend to bring starburst galaxies closer to the main sequence than what they are in reality, to the point where they can no longer be identified as such. Therefore, SFRs measured only from a single $24\,\um$ or ALMA flux at $z=2$ should not be used to study starburst galaxies.

\subsubsection{Dust mass \label{SEC:mono_dust}}

We present similar figures for $\mdust$ in \rfig{FIG:pred_disp} (bottom left and right) for submillimeter bands. The most striking fact to take out of this plot is that no band provides a measurement of $\mdust$ at better than $0.1\,\dex$ \citep[as observed in the local Universe in][]{groves2015}. This uncertainty rises steadily with redshift for the SPIRE bands, as rest-wavelengths get closer to the peak of the dust emission. This is partly compensated by the increase of the dust temperature with redshift (\req{EQ:tdust_ms}), which shifts the peak toward shorter wavelengths. As a consequence we found that the $870\,\um$ or $1.1\,{\rm mm}$ bands suffer roughly equivalent uncertainties, with only a minimal increase with redshift. \cite{scoville2014} recommended using rest-wavelengths larger than $250\,\um$ to measure the dust mass, which should rule out the use of $870\,\um$ beyond $z=2.5$. However they have assumed a constant and relatively cold average dust temperature of $25\kelvin$: for $\left<\tdust\right>=42\kelvin$ as we observed at $z=4$, the equivalent wavelength limit (in terms of distance to the peak) becomes $145\,\um$, which corresponds to $725\,\um$ at $z=4$. This implies that high frequency ALMA bands can still be used to trace the dust mass at high redshifts with a similar accuracy as $500\,\um$ at $z<1$ (see also \rsecs{SEC:alma_mdust_lir} where this is developed further).

The systematic bias on the dust mass of starburst galaxies is relatively constant with redshift, and most importantly never reaches zero. For these galaxies, we predict a systematic overestimation of the dust masses by a factor of at least $\rsb^{0.2}$, so a galaxy a factor five above the main sequence will have its dust mass overestimated by at least $40\%$. When the dust mass is converted to a gas mass (see \rsec{SEC:gas}), this implies that the gas fraction and depletion time of starbursts will be overestimated by a similar factor.

\subsection{Monochromatric conversion factors for ALMA and \jwst \label{SEC:conv}}

\begin{figure*}
    \centering
    \includegraphics[width=0.48\textwidth]{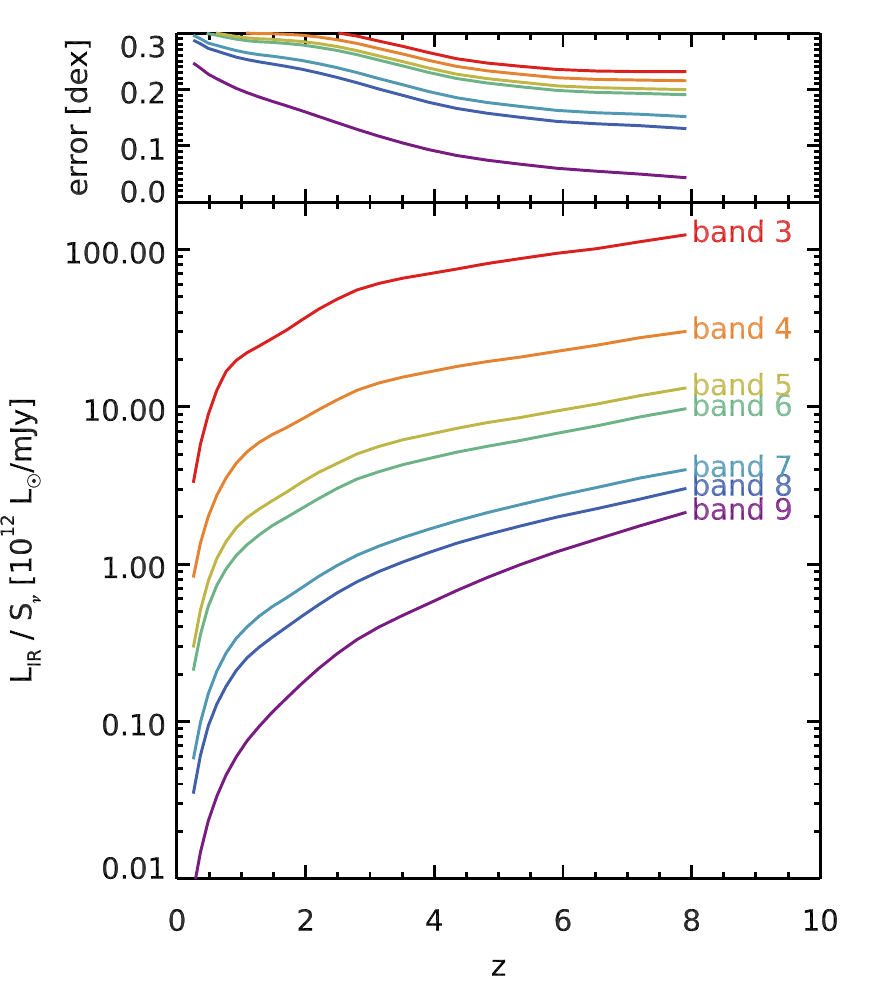}
    \includegraphics[width=0.48\textwidth]{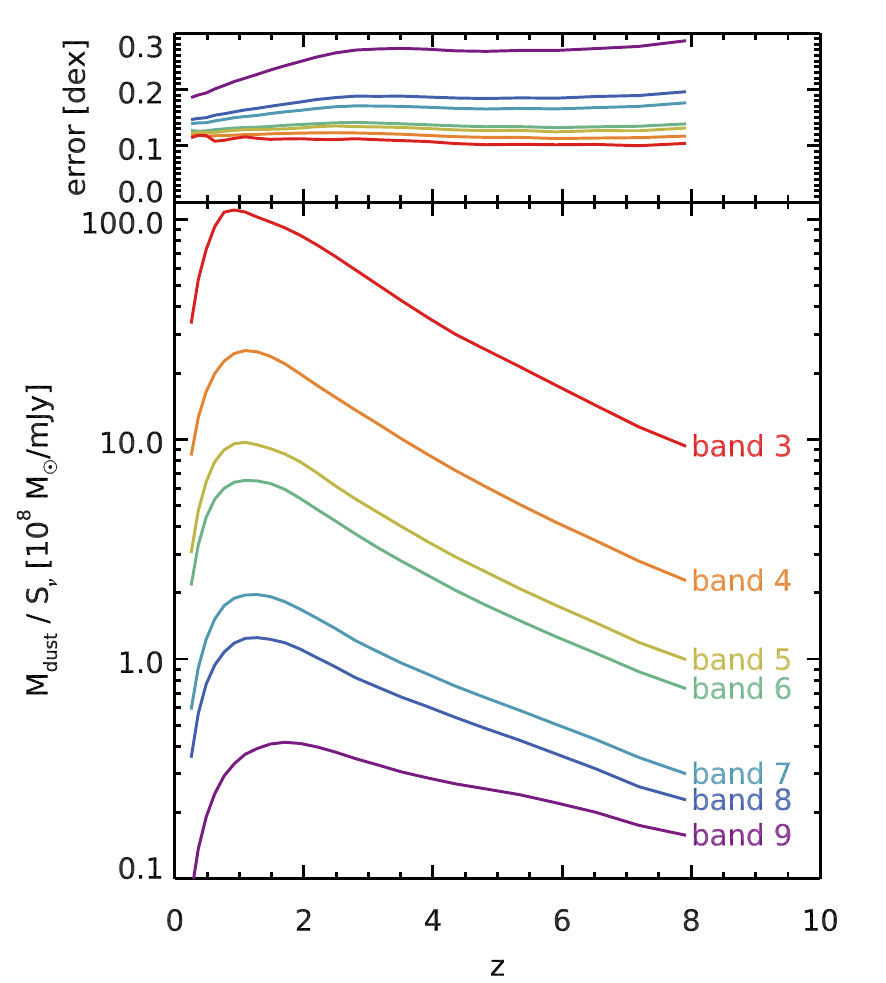}
    \caption{{\bf Left:} Conversion factor from observed flux (in mJy) to $\lir$ (in units of $12^{12}\,\lsun$). {\bf Right:} Conversion factor from observed flux (in mJy) to a dust mass (in units of $10^8\,\msun$). The top panels give the associated relative uncertainty on the conversion. These data are tabulated for easier access in \rtab{TAB:lconv_alma} and \rtab{TAB:dconv}, respectively. These factors are provided for all ALMA bands from band 8 to band 3, assuming the standard frequency.}
    \label{FIG:conv}
\end{figure*}

\begin{figure}
    \centering
    \includegraphics[width=0.48\textwidth]{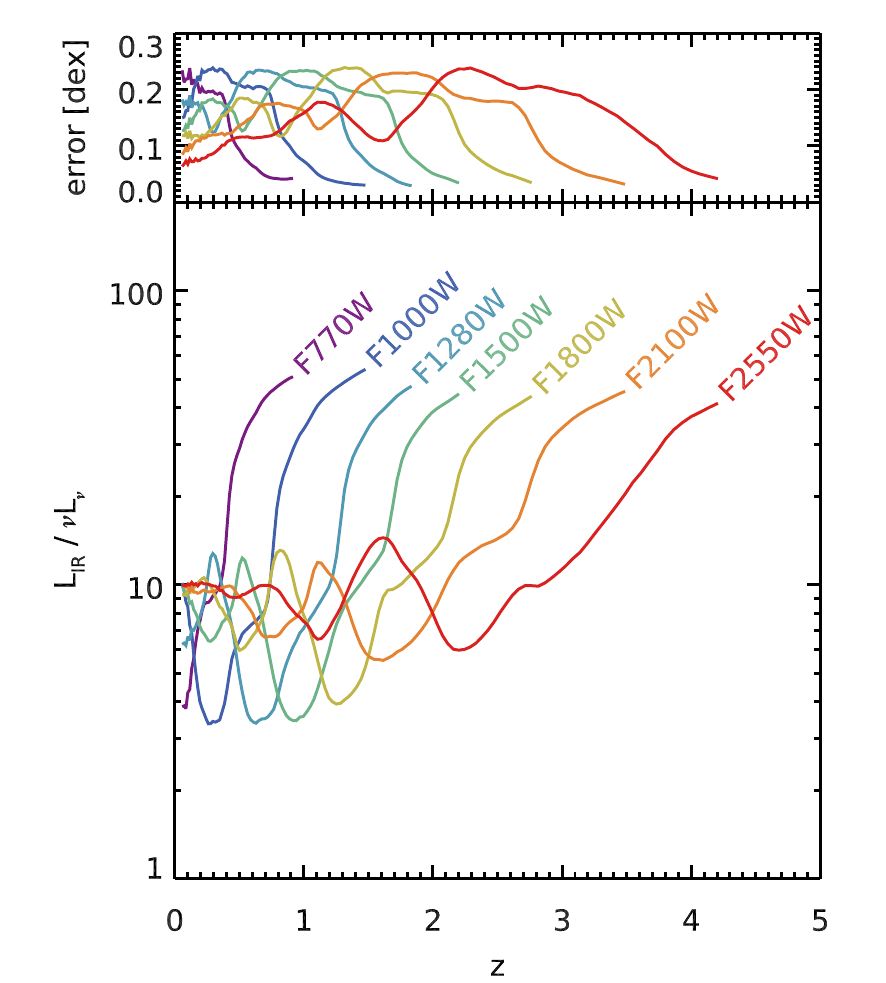}
    \caption{Conversion factor from observed monochromatic luminosity to $\lir$. These factors are provided for all \jwst MIRI bands, at all redshifts where the band probes the rest-frame $>3\,\um$ (i.e., where it may not be dominated by stellar continuum). The top panel gives the associated relative uncertainty on the conversion. These data are tabulated for easier access in \rtab{TAB:lconv}.}
    \label{FIG:conv_jwst}
\end{figure}

We provide numerical factors to convert an observed ALMA flux into a dust mass and infrared luminosity in \rtab{TAB:lconv_alma} and \rtab{TAB:dconv}, respectively, and conversion from a \jwst MIRI luminosity into $\lir$ in \rtab{TAB:lconv}. The later is truncated beyond the redshift where the \jwst bands probe the rest-frame $\lambda < 4\,\um$, where the stellar continuum starts to dominate the emission. These tables also include the conversion uncertainty (in dex), as shown on \rfig{FIG:pred_disp} for the \spitzer and \herschel bands. These data are also displayed in \rfigs{FIG:conv} and \ref{FIG:conv_jwst}.

It is clear from \rfig{FIG:conv_jwst} that the accuracy of a \jwst band in measuring $\lir$ will strongly depend on the redshift, as was already apparent for \spitzer $16$ and $24\,\um$. Depending on the redshift range of interest, it will therefore be more profitable to observe with one band or the other: for example, using F2550W instead of F1800W at $z=1.5$ can reduce the uncertainty on $\lir$ from $0.23$ to $0.12\,\dex$. This has to be considered together with the telescope's sensitivity in each band in determining the optimal observational setup (given, for example, that F2550W is expected to be seven time less sensitive than F1800W).

\subsubsection{Impact of redshift uncertainties for \jwst\label{SEC:delta_z}}

\begin{figure}
    \centering
    \includegraphics[width=0.48\textwidth]{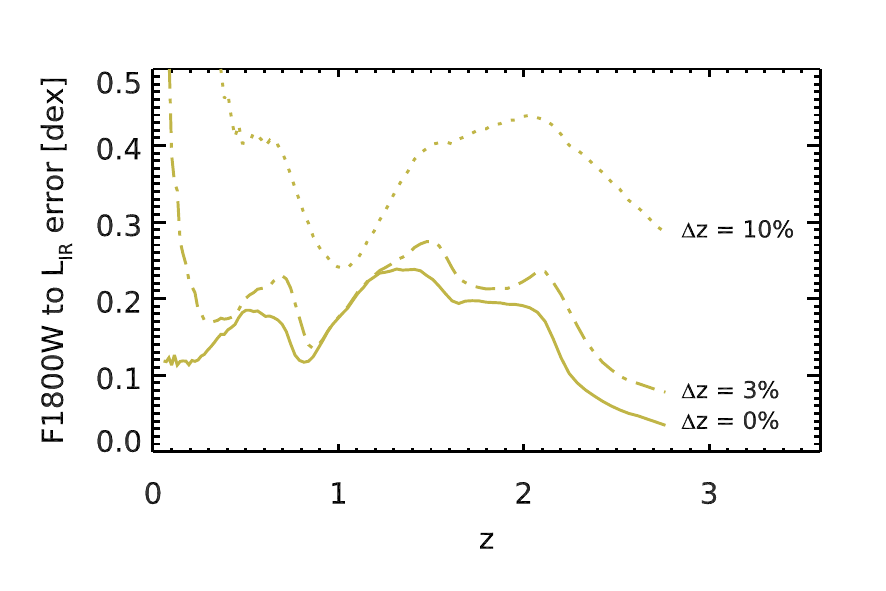}
    \caption{Uncertainty on the conversion of an observed \jwst F1800W luminosity into $\lir$. This uncertainty is shown with three cases: no uncertainty on the redshift (solid line), $3\%$ uncertainty (dot-dashed line), and $10\%$ uncertainty (dotted line).}
    \label{FIG:conv_jwst_dz}
\end{figure}

A potentially important consideration related to \jwst broadbands is the effect of redshift uncertainties. Because a large fraction of the flux in these bands comes from relatively narrow PAH emission lines, the conversion factor from flux to $\lir$ depends steeply on redshift as lines fall in and out of the bandpass. This can be precisely modeled if the spectroscopic redshift is known, but the case of photometric redshifts requires more care. To explore the impact of photometric redshift uncertainties, we have created two sets of ``observed redshifts'' for the galaxies in our mock catalog, obtained by randomly perturbing the true redshift with a Gaussian distribution of width $\Delta z$. We picked $\Delta z/(1+z)=3\%$ and $10\%$; $3\%$ is a typical (if not a conservative) value of the uncertainty in deep fields \citep{muzzin2013-a,skelton2014,pannella2015,straatman2016}, while $10\%$ is a more extreme value which applies only to rare outliers.

The result is shown in \rfig{FIG:conv_jwst_dz}, and cannot be described as a simple increase in quadrature from the case with no redshift uncertainty. In the case of $\Delta z/(1+z)=3\%$, the impact of the redshift uncertainty is null at $0.9 < z < 1.2$, and shows a maximal increase (in quadrature) of $0.16\,\dex$ at $z=0.7$, $1.5$ and $2.2$. This has the net effect of shifting the domains of best and worst accuracy toward slightly higher redshifts. This can be explained through the so-called ``negative $k$-correction'': when increasing the redshift, the decrease in flux caused by the larger distance can be compensated, partly or fully, if the intrinsic flux is higher at shorter rest-wavelengths. This happens when the broadband filter falls on the long-wavelength side of a PAH line. But globally, the uncertainty on the conversion is not dramatically increased, and the impact of the redshift uncertainty can be mostly ignored. In contrast, when $\Delta z/(1+z)=10\%$ the situation is much worse, with a minimal uncertainty of $0.25\,\dex$ around $z=1$, and $0.4\,\dex$ at other redshifts (i.e., essentially not a measurement). This highlights the sensitivity of the PAH region to redshift outliers, and suggests that multiple MIRI bands may be required to properly characterize the galaxies with the most uncertain redshifts (e.g., extremely dusty starbursts).

\subsubsection{ALMA: dust masses or infrared luminosities? \label{SEC:alma_mdust_lir}}

The dust mass and the infrared luminosity are the two main quantities one can measure from a set of ALMA fluxes; typically to infer gas masses and star formation rates, respectively. However, the conversion from an ALMA flux to either of these quantities depends on other factors, in particular on the dust temperature (see \req{EQ:lir_mdust}). The most secure approach would thus be to fit for the dust temperature, and then derive $\lir$ and $\mdust$. But when only a single flux measurement is available, the question arises: does this photometric point measure $\lir$, or $\mdust$? Because both depend on the dust temperature, the correct answer is ``neither'', which is not so helpful. However it is clear that one quantity will be constrained better than the other depending on the observed frequency (see previous sections and \rfig{FIG:pred_disp}): one expects high frequency bands to better trace $\lir$, and low frequency bands to better trace $\mdust$.

We quantified this question in two ways using our mock catalog. We first determined, for each band, the redshift ranges in which $\lir$ or $\mdust$ are measured at better than $0.2\,\dex$. We found that $\mdust$ is measured at better than $0.2\,\dex$ at all redshifts, and for all bands except band 9 ($440\,\um$, which is always worse than $0.2\,\dex$). In contrast, we found that $\lir$ can only be measured at this level of accuracy in band 9, 8, 7 and 6 and at $z>1$, $3.2$, $3.8$ and $5.7$, respectively. Therefore there are domains where either $\lir$ or $\mdust$ can be measured at better than $0.2\,\dex$ from the same ALMA measurement. We note however that only one of the two quantities can really be ``measured'', the other is then fully determined by the assumed $\tdust$.

Alternatively, for each redshift and band, we determined which of $\lir$ or $\mdust$ is constrained with the lowest uncertainty. The answer is always $\mdust$, except for band 9, 8 and 7 at $z>0.8$, $3.6$ and $5.7$, respectively, where $\lir$ takes over. Therefore, while band 7 ($870\,\um$) can be used starting from $z=3.8$ to measure $\lir$, it is only at $z>5.7$ that it traces $\lir$ better than it traces $\mdust$.

These results depend strongly on the evolution of the average dust temperature with redshift. If it had remained constant at the $z=0$ value, high frequency ALMA bands would have traced $\lir$ better at lower redshifts. This highlights that a proper knowledge of the dust temperature is crucial to interpret sub-millimeter fluxes, especially in cases where a single band is used at multiple redshifts to determine, for example, luminosity functions or cosmic $\sfr$ densities, or to build scaling laws.

\subsection{Gas masses from dust masses \label{SEC:gas}}

While the dust mass in itself is only of moderate interest, it is often used as a proxy for determining gas masses, though the assumption of a gas-to-dust ratio. In \cite{schreiber2016}, we have derived gas-to-dust ratios appropriate for our library: since systematic uncertainties on the dust masses are significant (see \rsec{SEC:amorphous}), it is crucial to use a consistent calibration of gas-to-dust ratios, derived using the same dust model. We obtained
\begin{align}
\frac{\mgas}{\mdust} = (155 \pm 23)\times\frac{Z_{\sun}}{Z}\,, \label{EQ:gdr}
\end{align}
where $\mgas$ is the total mass of atomic and molecular hydrogen, including helium, and $Z$ is the metallicity. This relation assumes that the metallicity is inferred from the oxygen abundance $12 + \log({\rm O}/{\rm H})$, measured using the \cite{pettini2004} calibration (see \citealt{magdis2012}), a solar metallicity of $Z_\sun=0.0134$ and a solar oxygen abundance of $8.73$ \citep{asplund2009}. The Pettini \& Pagel calibration was used by \cite{mannucci2010} to build the Fundamental Metallicity Relation (FMR), therefore the above formula can be applied directly to metallicities estimated using the FMR.

The uncertainty in the above formula is only statistical. Using galaxies in the HRS with well measured dust masses and independent measurement of gas masses from CO and H\textsc{i}, we determined that gas masses derived from dust masses were accurate at the level of $0.2\,\dex$. We emphasize that this gas-to-dust ratio was empirically calibrated using measured dust masses of nearby galaxies, and therefore the value of this ratio depends entirely on the model used to infer the dust masses. Using the DL07 model, for example, would produce higher dust masses by a factor two, hence the gas-to-dust ratio would have to be reduced by the same factor to remain consistent. Ultimately, the gas masses derived from dust masses of any model should be the same, provided the gas-to-dust ratios are properly calibrated.

\section{Conclusions}

We have introduced a new library of infrared SEDs, publicly available on-line, with three degrees of freedom: the dust mass or infrared luminosity, the dust temperature $\tdust$, and the mid-to-total infrared color $\ireight\equiv\lir/\leight$.

Using this library, we fit stacked SEDs of complete samples of main sequence galaxies in the CANDELS fields and derived the redshift evolution of the average $\tdust$ and $\ireight$, recovering and extending the trends previously identified in the literature. We observed that the most massive galaxies ($\mstar > 10^{11}\,\msun$) at $z<1$ have a reduced $\tdust$, sign of a reduced star formation efficiency, and found that low mass galaxies ($\mstar < 10^{10}\,\msun$) have an increased $\ireight$, probably because of their lower metallicities. Aside from these two mass domains, we found the infrared SED of the stacked galaxies only depend on the redshift, confirming the existence of a universal dust SED for main sequence galaxies at each epoch of the Universe.

We then used our new library to model galaxies individually detected in the \herschel images to determine how $\tdust$ and $\ireight$ vary for galaxies located above the main sequence, that is, the starbursts, and measure for the first time the scatter of both quantities. We recovered previous claims of a positive correlation of $\tdust$ and $\ireight$ with the offset from the main sequence. Both trends hint that the star forming regions in starbursts are more compact than in the typical main sequence galaxy. We observed a low residual intrinsic scatter of $12\%$ in $\tdust$ and $0.18\,\dex$ for $\ireight$, confirming that most of the observed variations of these two parameters are captured by the relations we derived with redshift and offset from the main sequence.

We have implemented these relations and scatters in the Empirical Galaxy Generator (EGG) to predict the accuracy of monochromatic measurements of $\lir$ and $\mdust$, as provided now by \spitzer MIPS $24\,\um$ and ALMA, and in the future with \jwst. We found that $\lir$ is best measured by wavelengths close to the peak of the dust emission, with a minimal uncertainty of $0.05\,\dex$, while mid-IR bands such as those of \jwst have a typical uncertainty of $0.1$ to $0.25\,\dex$. The highest frequency bands of ALMA can also be used to determine $\lir$, with an uncertainty of $0.2\,\dex$ or less at $z>0.9$, $3.2$, $3.8$, and $5.7$ for bands 9, 8, 7, and 6, respectively. Using randomly perturbed redshifts, we found these values to be only moderately increased in case of a typical redshift uncertainty of $3\%$. When measuring flux leftward of the peak (in wavelength), $\lir$ is only barely biased for starburst galaxies, however measurements rightward of the peak or using the rest-frame $8\,\um$ will underestimate the $\lir$ of starburst galaxies to the point of artificially bringing them back within the upper envelope of the main sequence.

Using the same mock catalog, we determined that the dust masses are best determined from the longest wavelength bands with an uncertainty of less than $0.15\,\dex$. High-frequency ALMA bands such as band $8$ and $7$ can also be used with an uncertainty of $0.2\,\dex$, however these bands will also tend to overestimate the dust (and gas) masses of starburst galaxies.

Finally, we tabulated the coefficients to convert observed ALMA fluxes into $\mdust$ and $\lir$ and \jwst luminosities into $\lir$, and provided estimates of the uncertainty associated to this conversion.

These results and our library can be used immediately to interpret the many observations in the ALMA archive, which most often consist of a single measurement per galaxy. Furthermore, we expect this will also be most useful for future proposals, either for ALMA or \jwst, by providing accurate predictions for the expected flux range of individual galaxies at various epochs.

\begin{acknowledgements}

The authors want to thank the anonymous referee for their comments that improved the consistency and overall quality of this paper.

Most of the numerical analysis conducted in this work have been performed using {\tt phy++}, a free and open source C++ library for fast and robust numerical astrophysics (\hlink{cschreib.github.io/phypp/}).

This work is based on observations taken by the CANDELS Multi-Cycle Treasury Program with the NASA/ESA \hst, which is operated by the Association of Universities for Research in Astronomy, Inc., under NASA contract NAS5-26555.

This research was supported by the French Agence Nationale de la Recherche (ANR) project ANR-09-BLAN-0224 and by the European Commission through the FP7 SPACE project ASTRODEEP (Ref.No: 312725).
\end{acknowledgements}

\bibliographystyle{aa}
\bibliography{../bbib/full}

\appendix

\section{Tabulated conversion factors for $\lir$ and $\mdust$}

\begin{table*}
\centering
\caption{Conversion from flux to infrared luminosity (in $10^{12}\,\lsun/\mJy$) and uncertainty (in dex). \label{TAB:lconv_alma}}
\begin{tabular}{cccccccc}
\hline\hline \\[-2.5mm]
$z$ & band 9 & band 8 & band 7 & band 6 & band 5 & band 4 & band 3 \\
    & 678 GHz & 404 GHz & 343 GHz & 229 GHz & 202 GHz & 149 GHz & 96.3 GHz \\ \hline \\[-2.5mm]
0.25 & 0.008 (0.25) & 0.035 (0.29) & 0.057 (0.30) & 0.210 (0.31) & 0.295 (0.31) & 0.821 (0.32) & 3.290 (0.33) \\
0.37 & 0.015 (0.24) & 0.061 (0.28) & 0.099 (0.29) & 0.358 (0.31) & 0.512 (0.31) & 1.364 (0.32) & 5.797 (0.32) \\
0.49 & 0.023 (0.23) & 0.094 (0.27) & 0.150 (0.28) & 0.539 (0.30) & 0.780 (0.30) & 2.009 (0.31) & 8.936 (0.32) \\
0.62 & 0.033 (0.22) & 0.129 (0.27) & 0.209 (0.28) & 0.734 (0.30) & 1.087 (0.30) & 2.735 (0.31) & 12.75 (0.31) \\
0.76 & 0.046 (0.21) & 0.167 (0.26) & 0.272 (0.27) & 0.931 (0.29) & 1.392 (0.30) & 3.530 (0.30) & 16.79 (0.31) \\
0.92 & 0.059 (0.20) & 0.211 (0.26) & 0.337 (0.27) & 1.136 (0.29) & 1.701 (0.29) & 4.379 (0.30) & 19.72 (0.31) \\
1.09 & 0.075 (0.19) & 0.254 (0.25) & 0.399 (0.26) & 1.330 (0.29) & 1.978 (0.29) & 5.182 (0.30) & 22.12 (0.31) \\
1.28 & 0.093 (0.19) & 0.297 (0.25) & 0.466 (0.26) & 1.532 (0.28) & 2.238 (0.29) & 5.937 (0.30) & 24.36 (0.31) \\
1.48 & 0.115 (0.18) & 0.343 (0.25) & 0.536 (0.26) & 1.757 (0.28) & 2.530 (0.29) & 6.657 (0.30) & 27.21 (0.31) \\
1.70 & 0.140 (0.17) & 0.397 (0.24) & 0.610 (0.26) & 1.979 (0.28) & 2.865 (0.29) & 7.378 (0.30) & 30.70 (0.31) \\
1.94 & 0.174 (0.16) & 0.466 (0.24) & 0.708 (0.25) & 2.261 (0.28) & 3.319 (0.29) & 8.346 (0.30) & 35.68 (0.31) \\
2.20 & 0.217 (0.15) & 0.552 (0.23) & 0.836 (0.25) & 2.615 (0.27) & 3.832 (0.28) & 9.595 (0.29) & 41.80 (0.30) \\
2.49 & 0.270 (0.14) & 0.656 (0.22) & 0.980 (0.24) & 3.026 (0.27) & 4.367 (0.28) & 11.06 (0.29) & 48.33 (0.30) \\
2.80 & 0.332 (0.13) & 0.772 (0.21) & 1.140 (0.23) & 3.477 (0.26) & 5.018 (0.27) & 12.74 (0.28) & 55.34 (0.29) \\
3.14 & 0.398 (0.12) & 0.898 (0.20) & 1.302 (0.22) & 3.867 (0.25) & 5.604 (0.26) & 14.18 (0.27) & 60.88 (0.29) \\
3.51 & 0.472 (0.10) & 1.031 (0.19) & 1.473 (0.21) & 4.284 (0.24) & 6.164 (0.25) & 15.45 (0.26) & 65.69 (0.28) \\
3.91 & 0.561 (0.09) & 1.181 (0.18) & 1.662 (0.20) & 4.687 (0.23) & 6.674 (0.24) & 16.64 (0.25) & 69.96 (0.27) \\
4.35 & 0.678 (0.08) & 1.356 (0.17) & 1.881 (0.19) & 5.141 (0.22) & 7.290 (0.23) & 18.05 (0.24) & 75.17 (0.25) \\
4.82 & 0.822 (0.07) & 1.540 (0.16) & 2.125 (0.18) & 5.608 (0.21) & 7.931 (0.22) & 19.39 (0.23) & 81.47 (0.25) \\
5.34 & 0.994 (0.07) & 1.748 (0.15) & 2.389 (0.17) & 6.117 (0.20) & 8.553 (0.21) & 20.70 (0.23) & 87.66 (0.24) \\
5.91 & 1.199 (0.06) & 1.989 (0.14) & 2.708 (0.16) & 6.781 (0.20) & 9.419 (0.21) & 22.53 (0.22) & 94.57 (0.24) \\
6.52 & 1.438 (0.05) & 2.249 (0.14) & 3.065 (0.16) & 7.553 (0.19) & 10.40 (0.20) & 24.61 (0.22) & 101.3 (0.23) \\
7.19 & 1.748 (0.05) & 2.587 (0.14) & 3.519 (0.16) & 8.614 (0.19) & 11.77 (0.20) & 27.49 (0.22) & 111.9 (0.23) \\
7.92 & 2.135 (0.04) & 3.032 (0.13) & 3.990 (0.15) & 9.739 (0.19) & 13.20 (0.20) & 30.18 (0.22) & 124.3 (0.23) \\
\hline
\end{tabular}
\end{table*}

\begin{table*}
\centering
\caption{Conversion from flux to dust mass (in $10^8\,\msun/\mJy$) and uncertainty (in dex). \label{TAB:dconv}}
\begin{tabular}{cccccccc}
\hline\hline \\[-2.5mm]
$z$ & band 9 & band 8 & band 7 & band 6 & band 5 & band 4 & band 3 \\
    & 678 GHz & 404 GHz & 343 GHz & 229 GHz & 202 GHz & 149 GHz & 96.3 GHz \\ \hline \\[-2.5mm]
0.25 & 0.085 (0.19) & 0.355 (0.15) & 0.588 (0.14) & 2.157 (0.13) & 3.036 (0.12) & 8.430 (0.12) & 33.62 (0.11) \\
0.37 & 0.137 (0.19) & 0.566 (0.15) & 0.914 (0.14) & 3.311 (0.13) & 4.741 (0.12) & 12.61 (0.12) & 53.35 (0.12) \\
0.49 & 0.191 (0.19) & 0.768 (0.15) & 1.230 (0.14) & 4.414 (0.13) & 6.398 (0.12) & 16.47 (0.12) & 73.09 (0.12) \\
0.62 & 0.244 (0.20) & 0.939 (0.15) & 1.520 (0.14) & 5.336 (0.13) & 7.918 (0.12) & 19.90 (0.12) & 92.96 (0.11) \\
0.76 & 0.294 (0.21) & 1.075 (0.16) & 1.749 (0.15) & 5.987 (0.13) & 8.951 (0.13) & 22.69 (0.12) & 108.0 (0.11) \\
0.92 & 0.334 (0.21) & 1.182 (0.16) & 1.891 (0.15) & 6.371 (0.13) & 9.545 (0.13) & 24.56 (0.12) & 110.6 (0.11) \\
1.09 & 0.369 (0.22) & 1.241 (0.16) & 1.954 (0.15) & 6.504 (0.13) & 9.679 (0.13) & 25.34 (0.12) & 108.2 (0.12) \\
1.28 & 0.391 (0.23) & 1.250 (0.17) & 1.966 (0.15) & 6.458 (0.13) & 9.430 (0.13) & 25.01 (0.12) & 102.6 (0.11) \\
1.48 & 0.410 (0.23) & 1.228 (0.17) & 1.922 (0.16) & 6.296 (0.13) & 9.067 (0.13) & 23.85 (0.12) & 97.50 (0.11) \\
1.70 & 0.418 (0.24) & 1.187 (0.17) & 1.824 (0.16) & 5.921 (0.14) & 8.579 (0.13) & 22.09 (0.12) & 91.84 (0.11) \\
1.94 & 0.413 (0.25) & 1.109 (0.18) & 1.685 (0.16) & 5.380 (0.14) & 7.893 (0.13) & 19.85 (0.12) & 84.81 (0.11) \\
2.20 & 0.398 (0.26) & 1.013 (0.18) & 1.533 (0.17) & 4.797 (0.14) & 7.029 (0.13) & 17.59 (0.12) & 76.64 (0.11) \\
2.49 & 0.377 (0.26) & 0.920 (0.19) & 1.375 (0.17) & 4.247 (0.14) & 6.131 (0.14) & 15.53 (0.12) & 67.83 (0.11) \\
2.80 & 0.352 (0.27) & 0.822 (0.19) & 1.213 (0.17) & 3.706 (0.14) & 5.347 (0.13) & 13.58 (0.12) & 58.99 (0.11) \\
3.14 & 0.330 (0.27) & 0.746 (0.19) & 1.083 (0.17) & 3.219 (0.14) & 4.663 (0.13) & 11.80 (0.12) & 50.66 (0.11) \\
3.51 & 0.307 (0.27) & 0.672 (0.19) & 0.960 (0.17) & 2.792 (0.14) & 4.017 (0.13) & 10.07 (0.12) & 42.83 (0.11) \\
3.91 & 0.288 (0.27) & 0.608 (0.19) & 0.856 (0.17) & 2.413 (0.14) & 3.434 (0.13) & 8.560 (0.12) & 35.99 (0.11) \\
4.35 & 0.270 (0.27) & 0.542 (0.18) & 0.753 (0.17) & 2.057 (0.14) & 2.917 (0.13) & 7.217 (0.12) & 30.07 (0.10) \\
4.82 & 0.256 (0.27) & 0.482 (0.18) & 0.665 (0.17) & 1.756 (0.13) & 2.487 (0.13) & 6.076 (0.12) & 25.53 (0.10) \\
5.34 & 0.241 (0.27) & 0.427 (0.18) & 0.584 (0.17) & 1.496 (0.13) & 2.091 (0.13) & 5.060 (0.11) & 21.43 (0.10) \\
5.91 & 0.221 (0.27) & 0.370 (0.18) & 0.504 (0.17) & 1.263 (0.13) & 1.753 (0.12) & 4.191 (0.11) & 17.60 (0.10) \\
6.52 & 0.201 (0.27) & 0.317 (0.19) & 0.431 (0.17) & 1.064 (0.13) & 1.463 (0.13) & 3.457 (0.11) & 14.27 (0.10) \\
7.19 & 0.175 (0.28) & 0.262 (0.19) & 0.356 (0.17) & 0.876 (0.13) & 1.190 (0.13) & 2.790 (0.11) & 11.38 (0.10) \\
7.92 & 0.158 (0.29) & 0.229 (0.20) & 0.300 (0.18) & 0.733 (0.14) & 0.993 (0.13) & 2.274 (0.12) & 9.301 (0.10) \\
\hline
\end{tabular}
\end{table*}

\begin{table*}
\centering
\caption{Conversion from monochromatic luminosity to $\lir$ and uncertainty (in dex). \label{TAB:lconv}}
\begin{tabular}{cccccccc}
\hline\hline \\[-2.5mm]
$z$ & F777W  & F1000W & F1280W & F1500W & F1800W & F2100W & F2550W \\
    & 7.66$\,\um$ & 9.97$\,\um$ & 12.8$\,\um$ & 15.1$\,\um$ & 18.0$\,\um$ & 20.8$\,\um$ & 25.4$\,\um$ \\ \hline \\[-2.5mm]
0.06 & 3.849 (0.23) & 10.15 (0.15) & 6.267 (0.18) & 9.897 (0.13) & 9.247 (0.12) & 9.947 (0.08) & 9.938 (0.06) \\
0.07 & 3.870 (0.21) & 9.508 (0.15) & 6.331 (0.17) & 9.568 (0.13) & 9.245 (0.12) & 9.969 (0.09) & 9.946 (0.07) \\
0.09 & 3.788 (0.22) & 8.741 (0.16) & 6.213 (0.18) & 9.206 (0.14) & 9.191 (0.12) & 9.822 (0.10) & 9.911 (0.07) \\
0.10 & 4.293 (0.22) & 8.415 (0.16) & 6.653 (0.17) & 9.256 (0.13) & 9.576 (0.11) & 9.973 (0.09) & 10.05 (0.07) \\
0.12 & 4.403 (0.24) & 7.280 (0.19) & 6.473 (0.19) & 8.640 (0.15) & 9.316 (0.13) & 9.520 (0.10) & 9.974 (0.08) \\
0.13 & 5.192 (0.21) & 6.863 (0.18) & 7.059 (0.17) & 8.741 (0.14) & 9.815 (0.11) & 9.734 (0.10) & 10.18 (0.07) \\
0.15 & 5.610 (0.22) & 5.802 (0.20) & 6.997 (0.18) & 8.233 (0.15) & 9.722 (0.12) & 9.427 (0.10) & 9.985 (0.07) \\
0.16 & 6.333 (0.21) & 5.074 (0.21) & 7.216 (0.17) & 8.006 (0.15) & 9.944 (0.12) & 9.435 (0.11) & 10.08 (0.08) \\
0.18 & 7.003 (0.21) & 4.469 (0.22) & 7.527 (0.18) & 7.697 (0.16) & 10.10 (0.12) & 9.349 (0.11) & 9.907 (0.07) \\
0.19 & 7.470 (0.19) & 4.015 (0.21) & 7.727 (0.17) & 7.223 (0.16) & 10.31 (0.11) & 9.393 (0.11) & 10.09 (0.07) \\
0.21 & 8.125 (0.20) & 3.799 (0.23) & 8.214 (0.17) & 6.953 (0.18) & 10.48 (0.12) & 9.523 (0.11) & 10.16 (0.08) \\
0.23 & 8.563 (0.20) & 3.651 (0.23) & 8.744 (0.16) & 6.814 (0.18) & 10.56 (0.12) & 9.594 (0.11) & 10.06 (0.08) \\
0.24 & 8.684 (0.19) & 3.492 (0.23) & 9.610 (0.15) & 6.636 (0.18) & 10.33 (0.12) & 9.532 (0.11) & 10.06 (0.08) \\
0.26 & 8.694 (0.20) & 3.364 (0.23) & 10.98 (0.14) & 6.443 (0.18) & 10.08 (0.12) & 9.442 (0.12) & 10.03 (0.08) \\
0.28 & 8.859 (0.20) & 3.370 (0.24) & 12.36 (0.13) & 6.420 (0.18) & 9.884 (0.13) & 9.463 (0.12) & 9.975 (0.08) \\
0.29 & 9.153 (0.20) & 3.425 (0.24) & 12.76 (0.12) & 6.510 (0.18) & 9.566 (0.13) & 9.468 (0.12) & 9.910 (0.09) \\
0.31 & 9.309 (0.19) & 3.408 (0.23) & 12.52 (0.13) & 6.624 (0.18) & 9.136 (0.14) & 9.543 (0.12) & 9.892 (0.09) \\
0.33 & 9.584 (0.19) & 3.432 (0.23) & 11.79 (0.13) & 6.864 (0.18) & 8.754 (0.14) & 9.611 (0.12) & 9.768 (0.09) \\
0.35 & 9.891 (0.19) & 3.473 (0.24) & 10.69 (0.15) & 7.084 (0.18) & 8.404 (0.15) & 9.628 (0.12) & 9.571 (0.10) \\
0.37 & 10.67 (0.19) & 3.700 (0.24) & 9.885 (0.16) & 7.431 (0.18) & 8.283 (0.15) & 9.821 (0.12) & 9.548 (0.10) \\
0.39 & 12.06 (0.18) & 3.937 (0.23) & 9.077 (0.16) & 7.565 (0.17) & 7.962 (0.15) & 9.809 (0.12) & 9.333 (0.10) \\
0.40 & 15.62 (0.16) & 4.396 (0.23) & 8.450 (0.17) & 7.779 (0.17) & 7.724 (0.16) & 9.869 (0.12) & 9.214 (0.11) \\
0.42 & 20.09 (0.14) & 4.994 (0.22) & 7.885 (0.18) & 8.034 (0.17) & 7.530 (0.16) & 9.908 (0.12) & 9.086 (0.11) \\
0.44 & 23.49 (0.12) & 5.598 (0.21) & 7.199 (0.18) & 8.424 (0.16) & 7.195 (0.17) & 9.838 (0.12) & 9.066 (0.11) \\
0.46 & 25.81 (0.11) & 5.963 (0.21) & 6.410 (0.19) & 9.179 (0.16) & 6.702 (0.17) & 9.619 (0.13) & 9.046 (0.11) \\
0.48 & 27.55 (0.10) & 6.221 (0.21) & 5.542 (0.20) & 10.52 (0.14) & 6.199 (0.18) & 9.314 (0.13) & 9.057 (0.11) \\
0.50 & 29.15 (0.09) & 6.468 (0.21) & 4.832 (0.21) & 11.74 (0.13) & 5.977 (0.18) & 9.067 (0.14) & 9.072 (0.12) \\
0.52 & 31.14 (0.08) & 6.794 (0.21) & 4.326 (0.22) & 12.32 (0.13) & 6.040 (0.18) & 8.934 (0.14) & 9.240 (0.12) \\
0.54 & 32.91 (0.08) & 6.964 (0.20) & 3.914 (0.22) & 12.06 (0.13) & 6.138 (0.18) & 8.696 (0.14) & 9.282 (0.12) \\
0.56 & 34.50 (0.07) & 7.103 (0.21) & 3.628 (0.23) & 11.28 (0.14) & 6.315 (0.18) & 8.446 (0.15) & 9.398 (0.12) \\
0.59 & 35.85 (0.07) & 7.259 (0.20) & 3.483 (0.23) & 10.47 (0.15) & 6.516 (0.18) & 8.210 (0.15) & 9.518 (0.11) \\
0.61 & 37.09 (0.06) & 7.403 (0.20) & 3.412 (0.23) & 9.770 (0.15) & 6.724 (0.18) & 7.912 (0.15) & 9.660 (0.11) \\
0.63 & 38.44 (0.06) & 7.512 (0.20) & 3.380 (0.23) & 9.072 (0.16) & 6.924 (0.18) & 7.423 (0.16) & 9.740 (0.11) \\
0.65 & 40.17 (0.06) & 7.724 (0.21) & 3.451 (0.23) & 8.486 (0.17) & 7.254 (0.18) & 7.022 (0.17) & 9.896 (0.11) \\
0.67 & 41.80 (0.05) & 7.902 (0.20) & 3.493 (0.23) & 7.779 (0.18) & 7.592 (0.17) & 6.754 (0.17) & 9.939 (0.12) \\
0.70 & 43.04 (0.05) & 8.144 (0.20) & 3.500 (0.23) & 6.950 (0.19) & 8.054 (0.17) & 6.632 (0.17) & 9.942 (0.12) \\
0.72 & 44.20 (0.05) & 8.832 (0.20) & 3.548 (0.23) & 6.147 (0.20) & 9.048 (0.16) & 6.643 (0.17) & 9.950 (0.12) \\
0.74 & 45.26 (0.04) & 10.59 (0.19) & 3.644 (0.23) & 5.449 (0.20) & 10.63 (0.14) & 6.679 (0.17) & 9.915 (0.12) \\
0.77 & 46.12 (0.04) & 14.02 (0.17) & 3.772 (0.23) & 4.812 (0.21) & 12.01 (0.13) & 6.645 (0.17) & 9.765 (0.12) \\
0.79 & 47.07 (0.04) & 18.20 (0.14) & 4.078 (0.23) & 4.339 (0.22) & 12.83 (0.12) & 6.693 (0.18) & 9.637 (0.12) \\
0.81 & 47.94 (0.04) & 21.21 (0.12) & 4.497 (0.22) & 3.998 (0.22) & 13.10 (0.12) & 6.822 (0.17) & 9.401 (0.13) \\
0.84 & 48.82 (0.04) & 23.39 (0.11) & 4.988 (0.22) & 3.767 (0.23) & 12.99 (0.12) & 7.058 (0.17) & 9.062 (0.13) \\
0.86 & 49.63 (0.04) & 25.18 (0.11) & 5.402 (0.22) & 3.590 (0.23) & 12.50 (0.12) & 7.285 (0.17) & 8.688 (0.14) \\
0.89 & 50.37 (0.04) & 26.93 (0.10) & 5.769 (0.21) & 3.495 (0.23) & 11.67 (0.13) & 7.464 (0.17) & 8.390 (0.14) \\
0.92 & 50.96 (0.04) & 28.86 (0.09) & 6.138 (0.21) & 3.451 (0.23) & 10.51 (0.15) & 7.615 (0.17) & 8.151 (0.15) \\
0.94 &      --      & 30.67 (0.08) & 6.449 (0.21) & 3.449 (0.23) & 9.363 (0.16) & 7.756 (0.17) & 7.890 (0.15) \\
0.97 &      --      & 32.37 (0.08) & 6.851 (0.21) & 3.544 (0.23) & 8.623 (0.17) & 8.059 (0.17) & 7.757 (0.15) \\
0.99 &      --      & 33.63 (0.07) & 7.072 (0.20) & 3.562 (0.23) & 7.941 (0.17) & 8.323 (0.16) & 7.520 (0.16) \\
1.02 &      --      & 35.15 (0.07) & 7.403 (0.20) & 3.693 (0.23) & 7.451 (0.18) & 8.986 (0.16) & 7.351 (0.16) \\
1.05 &      --      & 36.91 (0.06) & 7.713 (0.20) & 3.852 (0.23) & 6.880 (0.19) & 10.05 (0.15) & 7.046 (0.17) \\
1.08 &      --      & 38.82 (0.06) & 8.033 (0.20) & 4.070 (0.23) & 6.217 (0.20) & 11.20 (0.14) & 6.692 (0.17) \\
1.11 &      --      & 40.68 (0.05) & 8.432 (0.20) & 4.417 (0.23) & 5.536 (0.21) & 11.90 (0.13) & 6.522 (0.18) \\
1.13 &      --      & 42.14 (0.05) & 8.845 (0.20) & 4.853 (0.22) & 4.927 (0.21) & 11.85 (0.13) & 6.560 (0.18) \\
1.16 &      --      & 43.44 (0.04) & 9.297 (0.20) & 5.358 (0.22) & 4.454 (0.22) & 11.39 (0.14) & 6.762 (0.18) \\
1.19 &      --      & 44.60 (0.04) & 9.808 (0.20) & 5.933 (0.22) & 4.125 (0.23) & 10.94 (0.14) & 7.054 (0.17) \\
1.22 &      --      & 45.76 (0.04) & 10.70 (0.19) & 6.664 (0.21) & 3.989 (0.23) & 10.62 (0.15) & 7.470 (0.17) \\
1.25 &      --      & 46.80 (0.04) & 12.50 (0.18) & 7.341 (0.21) & 3.928 (0.23) & 10.15 (0.15) & 7.867 (0.17) \\
1.28 &      --      & 47.81 (0.03) & 16.26 (0.16) & 7.901 (0.20) & 3.949 (0.24) & 9.554 (0.16) & 8.331 (0.16) \\
1.31 &      --      & 48.89 (0.03) & 21.01 (0.13) & 8.452 (0.20) & 4.057 (0.24) & 8.912 (0.17) & 8.990 (0.16) \\
1.35 &      --      & 49.92 (0.03) & 24.54 (0.11) & 8.886 (0.20) & 4.159 (0.24) & 8.112 (0.18) & 9.707 (0.15) \\
\hline
\end{tabular}
\end{table*}

\begin{table*}
\centering
\caption{\rtab{TAB:lconv} continued.}
\begin{tabular}{cccccccc}
\hline\hline \\[-2.5mm]
$z$ & F777W  & F1000W & F1280W & F1500W & F1800W & F2100W & F2550W \\
    & 7.66$\,\um$ & 9.97$\,\um$ & 12.8$\,\um$ & 15.1$\,\um$ & 18.0$\,\um$ & 20.8$\,\um$ & 25.4$\,\um$ \\ \hline \\[-2.5mm]
1.38 &      --      & 50.98 (0.03) & 27.08 (0.10) & 9.296 (0.20) & 4.306 (0.24) & 7.359 (0.19) & 10.47 (0.14) \\
1.41 &      --      & 52.01 (0.03) & 29.06 (0.09) & 9.701 (0.20) & 4.501 (0.24) & 6.730 (0.20) & 11.19 (0.14) \\
1.44 &      --      & 53.00 (0.03) & 30.87 (0.08) & 10.13 (0.19) & 4.778 (0.24) & 6.214 (0.21) & 11.91 (0.13) \\
1.48 &      --      & 54.01 (0.03) & 32.84 (0.07) & 10.66 (0.19) & 5.248 (0.23) & 5.865 (0.21) & 12.66 (0.12) \\
1.51 &      --      &      --      & 34.84 (0.07) & 11.18 (0.19) & 5.946 (0.22) & 5.668 (0.22) & 13.38 (0.12) \\
1.54 &      --      &      --      & 36.56 (0.06) & 11.69 (0.19) & 6.897 (0.22) & 5.570 (0.22) & 13.93 (0.11) \\
1.58 &      --      &      --      & 38.07 (0.06) & 12.29 (0.19) & 8.108 (0.21) & 5.564 (0.22) & 14.34 (0.11) \\
1.61 &      --      &      --      & 39.32 (0.05) & 13.03 (0.18) & 9.081 (0.20) & 5.526 (0.23) & 14.44 (0.11) \\
1.65 &      --      &      --      & 40.76 (0.05) & 14.81 (0.17) & 9.582 (0.19) & 5.620 (0.23) & 14.31 (0.11) \\
1.69 &      --      &      --      & 42.31 (0.04) & 18.25 (0.16) & 9.630 (0.20) & 5.734 (0.23) & 13.70 (0.12) \\
1.72 &      --      &      --      & 43.85 (0.04) & 22.77 (0.13) & 9.822 (0.20) & 5.904 (0.23) & 12.84 (0.13) \\
1.76 &      --      &      --      & 45.10 (0.03) & 26.51 (0.10) & 10.08 (0.20) & 6.021 (0.23) & 12.00 (0.14) \\
1.80 &      --      &      --      & 46.28 (0.03) & 29.31 (0.09) & 10.49 (0.20) & 6.208 (0.23) & 11.41 (0.15) \\
1.83 &      --      &      --      & 47.34 (0.03) & 31.37 (0.08) & 10.92 (0.19) & 6.407 (0.23) & 10.89 (0.16) \\
1.87 &      --      &      --      &      --      & 33.23 (0.07) & 11.38 (0.19) & 6.700 (0.23) & 10.35 (0.17) \\
1.91 &      --      &      --      &      --      & 35.04 (0.06) & 11.78 (0.19) & 7.038 (0.23) & 9.644 (0.18) \\
1.95 &      --      &      --      &      --      & 36.86 (0.06) & 12.22 (0.19) & 7.456 (0.22) & 8.833 (0.19) \\
1.99 &      --      &      --      &      --      & 38.50 (0.05) & 12.82 (0.19) & 7.995 (0.22) & 8.092 (0.20) \\
2.03 &      --      &      --      &      --      & 39.78 (0.05) & 13.47 (0.19) & 8.659 (0.22) & 7.342 (0.21) \\
2.07 &      --      &      --      &      --      & 40.85 (0.04) & 14.38 (0.18) & 9.493 (0.21) & 6.713 (0.22) \\
2.12 &      --      &      --      &      --      & 41.98 (0.04) & 16.29 (0.17) & 10.40 (0.20) & 6.292 (0.23) \\
2.16 &      --      &      --      &      --      & 43.21 (0.04) & 19.60 (0.15) & 11.18 (0.19) & 6.032 (0.23) \\
2.20 &      --      &      --      &      --      & 44.58 (0.03) & 23.68 (0.12) & 11.90 (0.19) & 5.988 (0.24) \\
2.24 &      --      &      --      &      --      &      --      & 26.92 (0.10) & 12.39 (0.18) & 6.027 (0.24) \\
2.29 &      --      &      --      &      --      &      --      & 29.31 (0.09) & 12.79 (0.18) & 6.137 (0.24) \\
2.33 &      --      &      --      &      --      &      --      & 31.12 (0.08) & 13.17 (0.18) & 6.330 (0.23) \\
2.38 &      --      &      --      &      --      &      --      & 32.80 (0.07) & 13.58 (0.18) & 6.634 (0.23) \\
2.43 &      --      &      --      &      --      &      --      & 34.44 (0.07) & 13.85 (0.18) & 6.995 (0.23) \\
2.47 &      --      &      --      &      --      &      --      & 36.05 (0.06) & 14.11 (0.18) & 7.460 (0.22) \\
2.52 &      --      &      --      &      --      &      --      & 37.51 (0.05) & 14.49 (0.18) & 8.036 (0.22) \\
2.57 &      --      &      --      &      --      &      --      & 38.74 (0.05) & 14.90 (0.18) & 8.615 (0.21) \\
2.62 &      --      &      --      &      --      &      --      & 39.87 (0.05) & 15.57 (0.18) & 9.198 (0.21) \\
2.66 &      --      &      --      &      --      &      --      & 41.08 (0.04) & 16.79 (0.17) & 9.663 (0.20) \\
2.71 &      --      &      --      &      --      &      --      & 42.44 (0.04) & 19.28 (0.15) & 9.927 (0.20) \\
2.76 &      --      &      --      &      --      &      --      & 43.74 (0.03) & 22.84 (0.13) & 9.923 (0.20) \\
2.82 &      --      &      --      &      --      &      --      &      --      & 26.47 (0.11) & 9.887 (0.21) \\
2.87 &      --      &      --      &      --      &      --      &      --      & 29.45 (0.09) & 10.15 (0.20) \\
2.92 &      --      &      --      &      --      &      --      &      --      & 31.60 (0.08) & 10.55 (0.20) \\
2.97 &      --      &      --      &      --      &      --      &      --      & 33.33 (0.07) & 11.03 (0.20) \\
3.03 &      --      &      --      &      --      &      --      &      --      & 34.99 (0.07) & 11.58 (0.19) \\
3.08 &      --      &      --      &      --      &      --      &      --      & 36.68 (0.06) & 12.25 (0.19) \\
3.14 &      --      &      --      &      --      &      --      &      --      & 38.13 (0.05) & 12.88 (0.19) \\
3.19 &      --      &      --      &      --      &      --      &      --      & 39.47 (0.05) & 13.80 (0.18) \\
3.25 &      --      &      --      &      --      &      --      &      --      & 40.67 (0.05) & 14.88 (0.18) \\
3.31 &      --      &      --      &      --      &      --      &      --      & 41.79 (0.04) & 16.04 (0.17) \\
3.37 &      --      &      --      &      --      &      --      &      --      & 43.00 (0.04) & 17.30 (0.16) \\
3.43 &      --      &      --      &      --      &      --      &      --      & 44.29 (0.04) & 18.73 (0.15) \\
3.49 &      --      &      --      &      --      &      --      &      --      & 45.48 (0.03) & 20.35 (0.14) \\
3.55 &      --      &      --      &      --      &      --      &      --      &      --      & 22.31 (0.13) \\
3.61 &      --      &      --      &      --      &      --      &      --      &      --      & 24.01 (0.12) \\
3.67 &      --      &      --      &      --      &      --      &      --      &      --      & 26.16 (0.11) \\
3.74 &      --      &      --      &      --      &      --      &      --      &      --      & 28.49 (0.10) \\
3.80 &      --      &      --      &      --      &      --      &      --      &      --      & 31.31 (0.08) \\
3.87 &      --      &      --      &      --      &      --      &      --      &      --      & 33.72 (0.07) \\
3.93 &      --      &      --      &      --      &      --      &      --      &      --      & 35.62 (0.06) \\
4.00 &      --      &      --      &      --      &      --      &      --      &      --      & 37.31 (0.06) \\
4.07 &      --      &      --      &      --      &      --      &      --      &      --      & 38.60 (0.05) \\
4.14 &      --      &      --      &      --      &      --      &      --      &      --      & 40.00 (0.05) \\
4.21 &      --      &      --      &      --      &      --      &      --      &      --      & 41.38 (0.04) \\
\hline
\end{tabular}
\end{table*}

\end{document}